\documentclass[sigconf,authorversion,nonacm]{acmart}
\AtBeginDocument{%
  }

\usepackage{algorithm}
\usepackage{algpseudocode}
\usepackage{subfigure}

\begin{document}

\title{ECO-COMM: An Ultra Low-Latency Event Camera based Optical Communication System}

\author{Chengling Xu} 
\email{cxu338@wisc.edu}
% \orcid{}
\affiliation{%
  \institution{University of Wisconsin-Madison}
  \city{Madison}
  \state{Wisconsin}
  \country{USA}
}

\author{Keigo Hirakawa}
\email{khirakawa1@udayton.edu}
% \orcid{}
\affiliation{%
  \institution{University of Dayton}
  \city{Dayton}
  \state{Ohio}
  \country{USA}
}

\author{Feng Ye} 
\email{feng.ye@wisc.edu}
% \orcid{}
\affiliation{%
  \institution{University of Wisconsin-Madison}
  \city{Madison}
  \state{Wisconsin}
  \country{USA}
}

\renewcommand{\shortauthors}{Xu et al.}

\begin{abstract}
Ultra-low-latency communication is critical for emerging emerging next‑generation applications such as XR, real-time control, and distributed sensing. We present ECO-COMM, an event-camera-based optical communication system for ultra-low-latency device association and lightweight information exchange. By exploiting the asynchronous sensing and microsecond-level temporal resolution of event cameras, ECO-COMM captures high-frequency optical signals without frame-based acquisition delays. We identify and analyze key hardware-induced challenges in event-camera communication, including timestamp inconsistency, readout contention, trailing effects, and the inevitable refractory period, and develop hardware-aware mitigation techniques to address them. Focusing on a single transmitter-receiver optical link, ECO-COMM establishes the feasibility of practical ultra-low-latency event-camera communication using commercially available hardware. A prototype implementation using an eight-LED transmitter and an off-the-shelf event camera achieves device association within $15~\mu$s, symbol latency as low as $100~\mu$s, and end-to-end latency below $8$~ms for 32-byte payloads at a $0.1\%$ bit error rate. ECO‑COMM establishes a practical and complementary communication paradigm for ultra‑low‑latency systems where responsiveness and temporal precision are paramount.
% Ultra‑low‑latency communication is a key requirement for emerging Next‑generation (NextG) applications such as immersive XR, real‑time collaborative control, and distributed sensing. This paper presents ECO‑COMM, an event‑camera‑based optical communication system designed to enable ultra low‑latency device association and lightweight information exchange. ECO‑COMM leverages the asynchronous sensing and microsecond ($\mu$s)‑level temporal resolution of event cameras to capture high‑frequency spatiotemporal light modulation, eliminating the latency overhead of frame‑based acquisition, handshake procedures, and centralized scheduling. We systematically analyze hardware‑induced challenges, including timestamp inconsistency, readout contention, trailing effects, and inevitable refractory period, and introduce targeted mitigation techniques to ensure robust operation. A complete prototype implementation using an eight‑LED transmitter and an off‑the‑shelf event camera demonstrates 10~$\mu$s-level device association and 100~$\mu$s-level symbol latency. Experimental results further show end‑to‑end transmission latency below 8~ms for 32‑byte payloads at 0.1\% bit error rates. ECO‑COMM establishes a practical and complementary communication paradigm for ultra‑low‑latency NextG systems where responsiveness and temporal precision are paramount.
\end{abstract}




% \received{TBD}
% \received[revised]{TBD}
% \received[accepted]{TBD}

\maketitle

% \vspace{-3mm}
\section{Introduction}

Next‑generation (NextG) wireless communication systems aim to deliver a transformative leap beyond conventional mobile connectivity~\cite{10812743}. In addition to improvements in throughput and reliability, a defining objective of NextG networks is ultra‑low end‑to‑end latency, targeting the sub‑millisecond (ms) and even microsecond ($\mu$s) regime~\cite{9762857}. Such extreme latency reduction is widely regarded as essential for emerging time‑critical applications that demand instantaneous device association, rapid context awareness, and tightly synchronized information exchange across multiple devices, such as interactive multi‑user virtual, augmented, and extended reality (VR/AR/XR) systems, cooperative robotics and industrial automation, remote teleoperation, vehicle‑to‑everything (V2X) coordination, and distributed sensing at the network edge~\cite{doka2025enablers, 10902529,qiao2021survey}. In these contexts, communication latency directly impacts control stability, user experience, and operational safety, imposing stringent requirements on both device association time and per‑packet transmission delay.
Despite significant advances in 5G and ongoing research toward 6G, achieving microsecond-level responsiveness for device discovery and time-critical information exchange remains challenging. In many emerging applications, the latency associated with device association, synchronization, and communication setup can be a significant component of the overall response time. These challenges motivate the exploration of complementary communication mechanisms capable of providing extremely fast association and lightweight information exchange for latency-sensitive applications.

In this paper, we propose ECO-COMM, an Event Camera based Optical COMMunication system designed for ultra low‑latency device discovery and lightweight information exchange. ECO-COMM leverages the asynchronous sensing paradigm and microsecond ($\mu$s)‑level temporal resolution of event cameras to capture high frequency optical signals with minimal delay. By encoding information directly into spatiotemporal light modulation patterns, ECO-COMM avoids frame‑based acquisition and centralized scheduling, enabling newly arriving devices to rapidly synchronize and exchange essential state information.
Event‑based vision sensors have recently attracted attention as an alternative sensing modality for high‑speed perception and communication. Event cameras asynchronously report brightness changes with microsecond temporal resolution and high dynamic range, making them well suited for fast optical signaling. Prior work has demonstrated their effectiveness in feature detection, tracking, marker‑based localization, and beacon‑assisted communication~\cite{10198747,10797688,9982016}. However, existing event‑camera‑based communication systems typically operate at low modulation frequencies or rely on frame‑level symbol structures, resulting in per‑symbol latencies on the order of tens of milliseconds (ms).
ECO-COMM explicitly addresses hardware‑induced latency, readout contention, and multi‑LED synchronization challenges inherent in event‑based optical sensing. Through a hardware‑aware system design and targeted mitigation strategies, ECO-COMM achieves 10 $\mu$s‑level device association (under $15~\mu$s) and 100 $\mu$s‑level symbol latency, for example $100~\mu$s per symbol at $0.5\%$ bit error rate (BER) and $166~\mu$s per symbol at zero BER. These capabilities enable millisecond‑scale information exchange while preserving robustness to sensor‑level artifacts. We implement a complete prototype system to validate ECO-COMM in practice. Using an eight‑LED transmitter panel and an off‑the‑shelf event camera, ECO-COMM achieves end‑to‑end latency of approximately $8$~ms for a 32‑byte transmission at $0.1\%$ BER.

This work focuses on a single transmitter-receiver optical link and investigates the fundamental feasibility of ultra-low-latency event-camera-based communication. Our primary objective is to characterize and mitigate hardware-induced timing effects that limit latency and reliability, thereby establishing a practical foundation for event-camera-based optical communication. Broader networking challenges, including multi-device coordination, medium access, and network-scale protocol design, are beyond the scope of this study and are left for future work.
By demonstrating the feasibility of ultra‑low‑latency optical communication using commercially available hardware, this work establishes ECO-COMM as a practical and complementary approach for NextG wireless systems, particularly in scenarios where responsiveness and temporal precision are more critical than sustained throughput.

% \vspace{-2mm}
\section{Related Work}

Ultra‑low‑latency communication has been a central objective of next‑generation wireless systems, particularly in the context of 5G and emerging 6G networks~\cite{10859271}. In 5G, ultra‑reliable low‑latency communication (URLLC) was introduced to support time‑critical applications such as industrial automation and V2X communication~\cite{10902529}. Techniques including grant‑free access~\cite{10980184}, flexible numerologies~\cite{11153011}, and mobile edge computing (MEC)~\cite{11098592} have been proposed to reduce air‑interface and end‑to‑end latency. Research toward 6G further explores sub‑terahertz and mmWave spectrum, reconfigurable intelligent surfaces (RIS), cell‑free massive MIMO, and AI‑driven resource orchestration to push latency limits even lower~\cite{11025837}. 
Despite these advances, RF‑based systems continue to face fundamental challenges, including spectrum contention in dense deployments~\cite{11058339}, control‑plane overhead, and non‑negligible association latency. These limitations hinder the ability of RF technologies to consistently guarantee ultra‑low latency in highly dynamic, multi‑user scenarios such as immersive XR and collaborative robotic control.

In parallel, optical wireless communication has emerged as a promising complementary paradigm. For example, technologies such as light‑fidelity (Li‑Fi)~\cite{linnartz2022led}, free‑space optical (FSO) communication~\cite{hayle2025high}, and visible light communication (VLC)~\cite{9968053} offer high data rates, low propagation delay, and immunity to electromagnetic interference. By exploiting the vast unlicensed optical spectrum and strong spatial confinement of light, optical systems provide advantages in spectral efficiency, physical‑layer security, and interference isolation, making them attractive for short‑range and interference‑sensitive environments. However, these technologies often rely on conventional frame‑based optical receivers, which incur non‑trivial latency due to exposure time, frame readout, and clock synchronization~\cite{9967970}.

Event‑based vision sensors have recently attracted attention as an alternative sensing modality that can mitigate some of these constraints and apply as optical communication systems. For example, this work~\cite{10797688} introduces the to use event cameras as receivers for data transmitted from standard digital display. However, the algorithm was designed for 60 Hz display, hence per-symbol latency is on the order of 16.7 ms or 33.3 ms with enhanced robustness. Therefore, it is not feasible to implement these algorithms for ultra low latency communications. The authors in~\cite{9982016} introduced an asynchronous optical communication system using event cameras, achieving data rates of up to 4 kbps indoors and 500 bps transmission over 100 m outdoors. However, per‑symbol latencies of these approaches are on the order of tens of ms or more. 

This gap motivates the development of ECO-COMM, a new optical communication frameworks that leverage event‑based sensing to enable 10 $\mu$s-level device association and 100 $\mu$s-level symbol latency for time‑critical information exchange.

% \vspace{-3mm}
\section{Limitations and Challenges on Accurate Timestamp Recording of Events}

Accurate event recording is critical for ultra low‑latency optical communication. In practice, however, limitations in pixel circuitry, analog bandwidth, and readout control introduce timestamp inaccuracies that complicate reliable protocol design. This section examines the most prominent sources of such inaccuracies under controlled LED illumination. Effects related to ambient lighting and camera parameter tuning are treated as part of standard event‑camera calibration and are not discussed. All results are obtained using the experimental setup (see Section~\ref{sec:setup}). While specific measurements vary across systems, the identified phenomena are generally applicable to event camera based optical communication.

% \vspace{-3mm}
\subsection{Inconsistency of Time Stamps}

Even under ideal illumination and constant ambient lighting conditions, event timestamps generated by individual pixels can exhibit noticeable inconsistency due to pixel front-end dynamics and non-instantaneous response. 
Each event pixel employs an analog front‑end composed of a photodiode, a logarithmic transimpedance (or log‑amplifier) stage, and a comparator. As a result, the pixel response to a step change in optical input is temporally smoothed by the finite bandwidth of the analog circuitry, rather than being instantaneous.
A convenient analytical approximation models the internal log‑intensity signal $\widetilde{L}_{x,y}(t)$ as a first‑order low‑pass filter of the true log‑intensity $L_{x,y}(t)$:
\begin{align}    
\lambda \frac{d \widetilde{L}_{x,y}(t)}{dt} + \widetilde{L}_{x,y}(t) = L_{x,y}(t),
\end{align}
where $\lambda>0$ denotes the analog time constant. For an LED step transition occurring at time $\tau$, the input and filtered responses are
\begin{align}\label{eq:edge}
L_{x,y}(t) &= B + A \, u(t-\tau),~
\widetilde{L}_{x,y}(t) = B + A \left(1 - e^{-(t-\tau)/\lambda}\right) u(t-\tau),
\end{align}
with step amplitude $A$, background level $B$, and unit step function $u(\cdot)$.
An event is triggered when the change in $\widetilde{L}_{x,y}(t)$ relative to the reset value $R=B+U$ exceeds the contrast threshold $C$, where $U\sim\mathrm{Unif}(0,C)$. Solving for the time‑to‑event $\Delta=t-\tau$ yields
\begin{align}
\widetilde{L}_{x,y}(t) - R = C \quad \Rightarrow \quad
A \left(1 - e^{-(t-\tau)/\lambda}\right) = C +U.
\end{align}
Solving for time-to-event $\Delta=t-\tau$, we have
\begin{align}
\Delta = - \lambda \log\left(1 - \frac{C+U}{A}\right) \approx 
\lambda \frac{C+U}{A}\sim \operatorname{Unif}\left(\lambda  \frac{C}{A},\lambda  \frac{2C}{A}\right),
\end{align}
assuming $A\gg C$. This result shows that event timestamps are delayed by a random offset proportional to the analog time constant $\lambda$ and threshold $C$, and inversely proportional to the signal amplitude $A$. In practice, device mismatch and operating conditions further complicate the delay distribution.

\vspace{-2mm}
\subsection{Inconsistent Sensor Readout Delay}

The asynchronous and sequential nature of event readout poses a fundamental challenge for high‑frequency event‑based sensing. Ideally, a state transition of an LED would trigger simultaneous responses from all observing pixels, yielding identical timestamps. In practical hardware, however, each recorded timestamp $t$ includes a readout‑dependent delay $\Delta t_{\text{readout}}$,
\begin{equation}
t = t_{0} + \Delta t_{\text{readout}},
\end{equation}
where $t_{0}$ denotes the true physical transition time. This delay is neither constant nor uniform across pixels.

\begin{figure}[ht!]
    \centering
    \vspace{-3mm}
    \subfigure[``ON'' events.]{
        \includegraphics[width=0.4\linewidth]{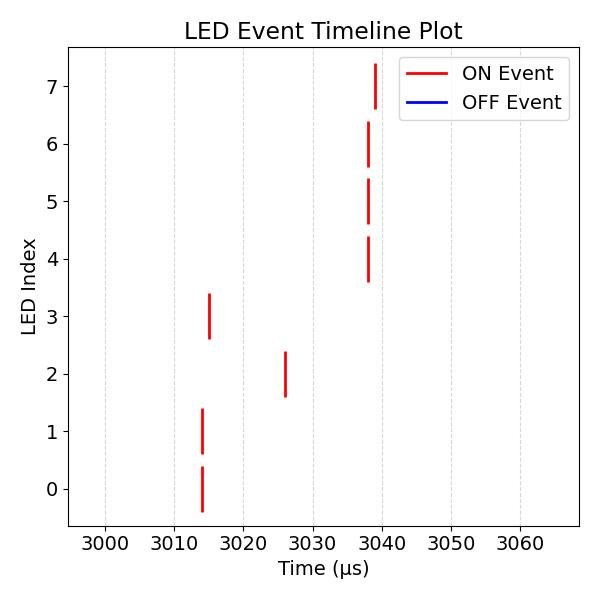}
        \label{fig:inter_led_delay_on}
    }%
    % \hfill
    \subfigure[``OFF'' events.]{
        \includegraphics[width=0.4\linewidth]{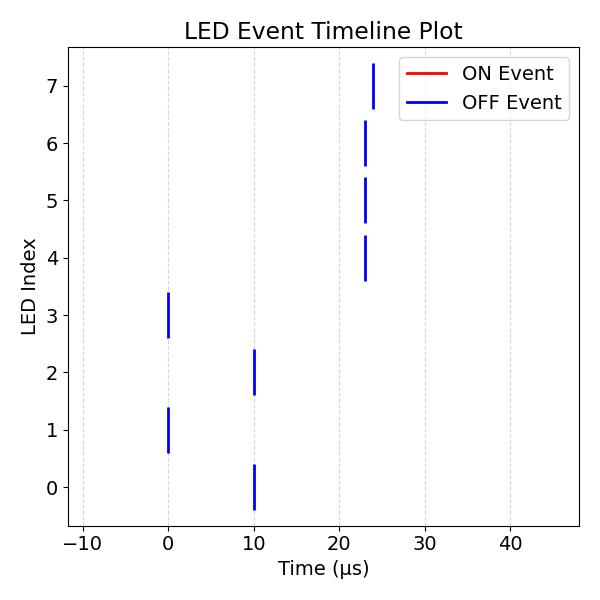}
        \label{fig:inter_led_delay_off}
    }
    \vspace{-3mm}
    \caption{Preliminary results on inter-LED readout delay.}
    \label{fig:inter_led_delay}
\end{figure}

As illustrated in Fig.~\ref{fig:inter_led_delay}, even when all eight LEDs are commanded to switch simultaneously, the recorded events are temporally dispersed. The readout delay is primarily influenced by three factors: (i) the camera–panel distance $d$, assuming a fixed focal length; (ii) the number of simultaneously active LEDs $N_{\mathrm{LED}}$; and (iii) the spatial coordinates $(x,y)$ of the activated pixel regions. These factors collectively determine the instantaneous event load on the readout bus and the servicing order of pixel requests.
The dominant contributor to readout latency is the instantaneous event volume. As $d$ decreases or $N_{\mathrm{LED}}$ increases, each LED activates more pixels, generating a higher burst of concurrent events. Since readout throughput is hardware‑limited, increased contention on the readout bus leads to longer and more variable per‑pixel delays.
In addition, the sensor’s deterministic readout order introduces spatially dependent timestamp skew. Pixels are typically serviced sequentially from the top‑left to the bottom‑right of the array, granting earlier timestamps to pixels with smaller $(x,y)$ coordinates. Consequently, LEDs whose images fall closer to the upper‑left region of the sensor produce earlier event timestamps than those in lower‑right regions, even if their physical transitions are simultaneous.
Preliminary measurements using an eight‑LED panel confirm this effect. As shown in Fig.~\ref{fig:inter_led_delay}, both ON and OFF transitions exhibit a temporal spread of approximately $25~\mu\mathrm{s}$ between the earliest and latest recorded events, highlighting the non‑negligible and systematic nature of inter‑LED readout delay.

% \vspace{-2mm}
\subsection{Trailing Effect}\label{sec:trailing}

\begin{figure}[b]
    \centering
    \vspace{-3mm}
    \includegraphics[width=0.99\linewidth]{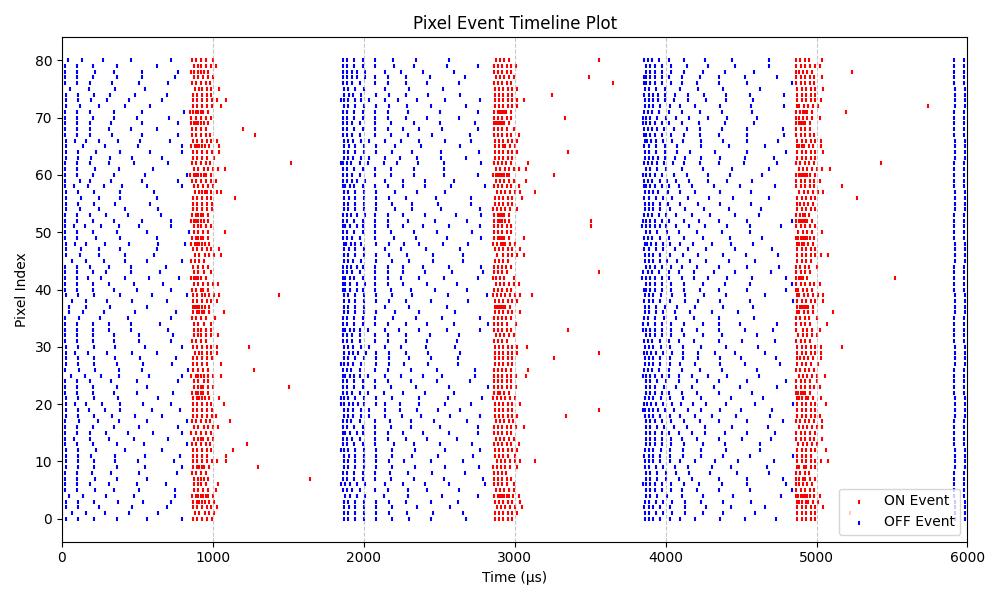}
    \vspace{-4mm}
    \caption{Preliminary results on trailing effects.}
    \label{fig:trailing_effect}
\end{figure}

A major challenge in high‑frequency event‑based sensing is the \emph{trailing effect}. Ideally, a state transition of an LED would trigger a single event at a well‑defined timestamp from each observing pixel. In practical hardware, however, a single LED transition often produces a temporally extended response. Rather than a single event, each photosensitive pixel emits a sequence of events over a finite duration, or until the next illumination change occurs.
This behavior arises from the interaction between the logarithmic response of the pixel analog front‑end (AFE) and its finite bandwidth. The photosensor converts the incident light intensity $I$ into a photoreceptor voltage $V_{\text{pr}}$ according to the logarithmic relationship $V_{\text{pr}} \propto \ln(I)$, which is stored on a sampling capacitor. An event is generated whenever the change in $V_{\text{pr}}$ exceeds a predefined contrast threshold $C_{\text{th}}$.
For a high‑contrast LED transition, such as OFF‑to‑ON, the resulting change $\Delta \ln(I)$ typically exceeds $C_{\text{th}}$ by a large margin. Consequently, the photoreceptor voltage rises from its initial value $V_{\text{pr}}(t_0)$ toward a steady‑state target $V_{\text{tgt}}$, crossing multiple threshold levels of the form $V_{\text{pr}}(t_0) + k C_{\text{th}}$. Each threshold crossing generates an event, leading to a burst of trailing events following a single physical transition.
The timing of these events is governed by the finite bandwidth of the AFE. The voltage across the sampling capacitor evolves according to
\begin{equation}
V_{\text{pr}}(t) = V_{\text{tgt}} - \bigl(V_{\text{tgt}} - V_{\text{pr}}(t_0)\bigr) e^{-(t - t_0)/\tau},
\end{equation}
where $\tau$ is the AFE time constant and relates to the cutoff frequency as $f_{\text{BW}} \approx 1/(2\pi\tau)$. The event firing rate is proportional to the voltage derivative,
\begin{equation}
\frac{d V_{\text{pr}}}{dt} =
\frac{V_{\text{tgt}} - V_{\text{pr}}(t_0)}{\tau}
e^{-(t - t_0)/\tau}.
\end{equation}
As $V_{\text{pr}}(t)$ approaches $V_{\text{tgt}}$, the derivative decays exponentially, resulting in progressively increasing temporal spacing between successive events. This behavior manifests as a trailing sequence with a decreasing firing rate, reflecting the fundamental bandwidth limitation of the pixel analog circuitry.

Fig.~\ref{fig:trailing_effect} shows a representative example from our preliminary experiments, in which six LED transitions are observed by 81 pixels. While these transitions ideally correspond to three pairs of OFF‑to‑ON and ON‑to‑OFF events, each nominal transition instead generates a burst of events across all activated pixels. In our measurements, trailing events can persist until the next LED transition. Besides, ON‑event trailing is less severe than OFF-event trailing.

\subsection{Inevitable Refractory Period}\label{sec:IRP}

Another critical temporal non‑ideality is the Inevitable Refractory Period (IRP). On an ideal sensor, a pixel should be ready to detect a new contrast change immediately after the completion of a reset pulse. However, physical event-based sensors exhibit a "blind" phase in this situation. This is different from the standard refractory period (RP), which is often a programmable parameter used to limit the maximum firing rate. The IRP is an intrinsic hardware bottleneck inherited from the settling dynamics of the pixel's analog front-end.
Specially in our prototype implementation, after a pixel transitions from a long-term quiescent state to an active firing state, there is a span of time (about 250 $\mu s$ in our test as described in section 6.9) that the pixel does not respond to any changes or generate any events. This issue originates in the feedback loop of the comparator circuit. As previously discussed, generating an event triggers a reset signal to update $V_{\text{ref}}$. However, in a physical circuit, this reset is not instantaneous. After a long period of inactivity, the internal nodes of the differential amplifier drift away from their nominal DC operating points due to parasitic capacitor voltage leakage. When the first event triggers a reset, since the amplifier operates with extremely low bias currents, it requires a non-neglible amount of time to recharge the internal capacitors and bring the amplifier circuit back to the nominal operating point. Therefore, the reference voltage $V_{\text{ref}}$ is unreliable, making the circuit not able to produce effective events. During this "warm-up" phase, the pixel’s effective contrast sensitivity $C_{\text{th}}$ is temporarily infinite, which is depicted as
\begin{equation}
C_{\text{eff}}(t) = \begin{cases} \infty, & t_0 < t < t_0 + t_{\text{IRP}}, \\ C_{\text{th}}, & t \ge t_0 + t_{\text{IRP}}. \end{cases}
\end{equation}
Unlike the trailing effect which gives extra unwanted events, the IRP introduces a systematic latency and potential loss of information for high frequency signals following a period of silence. Typically, $t_{\text{IRP}}$ resides in the order of hundreds of microseconds, introducing a fundamental limit on the temporal resolution to the sensor's \emph{cold-start} response.

\section{Mitigation Methods}\label{sec:method for robust control}

This section presents a set of mitigation strategies designed to address the limitations affecting accurate and robust event time stamping identified in the preceding analysis.

\subsection{Robustness to Temporal Jitter}
\label{sec:torlerance_window}

As discussed previously, the raw timestamps produced by the event camera are subject to stochastic temporal jitter introduced by sensor latency and asynchronous readout. Let the ideal timestamp for an event in the $m$‑th modulation period be defined as follows:
\begin{equation}
t_{m}^{\text{ideal}} = t_{\text{start}}^{\text{ideal}} + m \cdot T,
\end{equation}
where $t_{\text{start}}^{\text{ideal}}$ denotes the true start time of a frame and $T$ is the modulation period. The measured event timestamp is then modeled as follows:
\begin{equation}
t_{m} = t_{m}^{\text{ideal}} + \Delta t_m,
\end{equation}
with $\Delta t_m \ge 0$ representing sensor‑related latency. In practice, the receiver’s estimated start time also experiences a non‑negative delay,
\begin{equation}
t_{\text{start}} = t_{\text{start}}^{\text{ideal}} + \Delta t_{\text{start}},
\end{equation}
where $\Delta t_{\text{start}} \ge 0$. Substituting these expressions, the measured timestamp of the $m$‑th event relative to the receiver’s start reference can be written as
\begin{equation}
t_m = t_{\text{start}} + m \cdot T + \Delta t_m',
\end{equation}
where $\Delta t_m' = \Delta t_m - \Delta t_{\text{start}}$ is the relative synchronization error. Since $\Delta t_m'$ represents the difference between two independent, non‑negative stochastic variables, it may take either positive or negative values.

To tolerate this temporal uncertainty, we employ a temporal windowing strategy. Instead of sampling events at a single expected time instant, we consider all events falling within a tolerance window defined as $[m \cdot T - T/2,, m \cdot T + T/2)$. This widened window accommodates both early and delayed events caused by jitter and readout variability. By detecting LED state changes based on events within this interval, the system becomes robust to frame‑to‑frame variations in $\Delta t_m'$, thereby improving the reliability of symbol decoding under continuous high‑frequency modulation.

\subsection{Inter-LED Time Shifting for Readout Delay Compensation}\label{sec:readout_mitigation}

To alleviate readout saturation and reduce spatially dependent delays, we introduce a deterministic inter‑LED time‑shifting strategy at the transmitter. Rather than switching all LEDs simultaneously, each LED is assigned a fixed temporal offset according to its relative location. Assuming the LEDs are indexed from left to right, top to bottom, following the camera sensor readout sequence. Then, in each modulation period, the state transition of LED $k$ is delayed by
\begin{equation}
\Delta_k = k \cdot \Delta_{\text{base}},
\end{equation}
where $\Delta_{\text{base}}$ is a predefined constant, for example $2~\mu\text{s}$. As an illustration, LED 0 transitions at $0~\mu\text{s}$, LED 1 at $2~\mu\text{s}$, LED 2 at $4~\mu\text{s}$, and so forth.
This staggered activation spreads the event generation over time, significantly reducing instantaneous contention on the sensor readout bus. As a result, readout delays caused by bursty event traffic and spatial readout ordering are mitigated. The introduced temporal offsets are deterministic and are therefore compensated during subsequent decoding, preserving the correctness of the transmitted data.

% To reduce the sensor's readout rate limitations and the spatial-dependent delays, we implement a manual time-shifting strategy at the transmitter. Instead of triggering all LEDs in the panel simultaneously, we introduce a deterministic delay for each LED based on its index. Specifically, for each period, the state change for LED $k$ is delayed by:
% \begin{equation}
%     \delta_k = k \cdot \Delta_{\text{base}},
% \end{equation}
% where $\Delta_{base}$ is a predefined constant (e.g., $2 \mu s$). For instance, in a given period, LED 0 transitions at $0 \mu s$, LED 1 at $2 \mu s$, LED 2 at $4 \mu s$, and so on. By introducing these transitions, we effectively distribute the original simultaneous event generation over time, greatly easing the instantaneous event saturation of the sensor's readout bus. This controlled delay will be compensated in the following data processing.

\subsection{Event Aggregation and Noise Reduction}\label{sec:aggregation}

To mitigate the trailing effect, we adopt a two‑stage processing pipeline. In the first stage, trailing events are filtered on a per‑pixel basis. Since the current system encodes information solely through binary LED ON and OFF states, trailing events primarily reflect the AFE recovery dynamics rather than genuine state changes. Events that occur within a predefined trailing window following an initial transition are therefore classified as artifacts and discarded. In the second stage, we propose to aggregate filtered events at the LED level to infer state transitions as follows:
\begin{itemize}
    \item \textbf{Best Pixel (BP):} The trailing events of each pixel in ROIs is first removed. For each ROI, the time error of each pixel is pre-evaluated using the method in section 6.4 and the one with the lowest error is then used as the best pixel to represent the LED state change.

    \item \textbf{Sequential Polarity Accumulation (SPA):} Remove the trailing events of each pixel and put them in time order. Within each ROI, a state change is only recognized if $n$ continuous events of the same polarity are detected. The method is also written as S-n. 
    
    \item \textbf{Cluster (Clu):} Remove the trailing events of each pixel and put them in time order. Divide events into short window and apply majority voting for polarity. The result event is constructed with the mean timestamp and the majority polarity.
\end{itemize}
Because data transmission depends on the LED state rather than individual pixel behavior, this aggregation suppresses isolated noise events while preserving sensitivity to true LED transitions. The combined filtering and aggregation process can substantially improve robustness without sacrificing temporal responsiveness.

\subsection{IRP Mitigation}

A random start without synchronization between a transmitter and the receiver would encounter IRP no matter what. However, assuming the system operates in a static environment without mobility, then the transmitter may implement a pre-activitation strategy for IRP mitigation. Prior to the start of data transmission, the LEDs are turned on for several modulation periods. This initial activation stabilizes the pixel analog front‑end and allows the differencing circuitry to converge to a reliable operating point before meaningful data modulation begins. Note that more advanced initial activation, e.g., through AI-based prediction, is beyond the scope of this work.
Meanwhile, during inter‑frame idle periods, IRP can be mitigated by holding the LEDs in a constant ON state rather than being switched off. Allowing the LEDs to remain OFF during these pauses would reintroduce the IRP at the beginning of the next frame, increasing latency and temporal uncertainty. Experimental results show that alternating between ON and OFF states during idle periods prevents the pixels from maintaining a stable reference and can even prolong the IRP. By sustaining an ON baseline, the first transition of each new data frame is processed with minimal hardware‑induced delay, thereby preserving synchronization accuracy and symbol integrity.

\section{Ultra Low-Latency Communication Protocol}\label{sec:method for communiation}

\begin{figure}[ht!]
    \centering
    \vspace{-3mm}
    \includegraphics[width=0.9\linewidth]{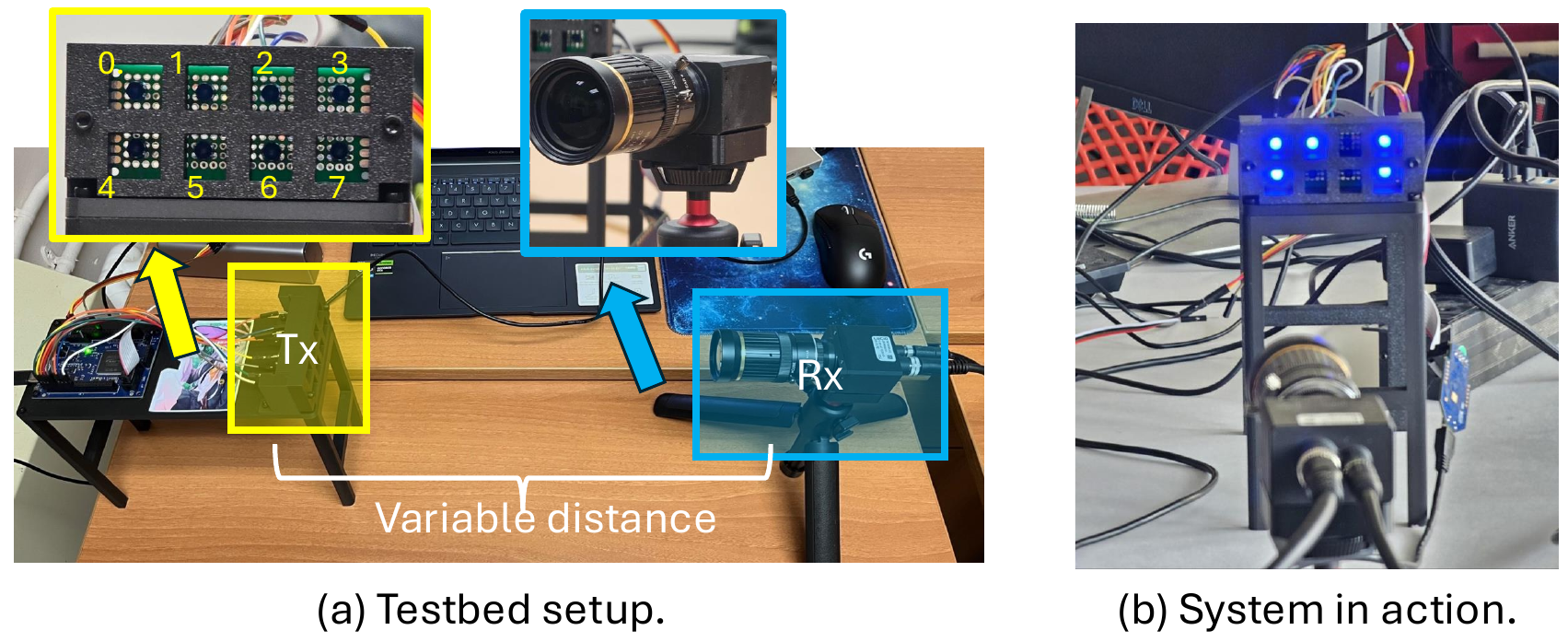}
    \vspace{-3mm}
    \caption{Prototype of ECO-COMM system.}
    \label{fig:testbed}
     \vspace{-3mm}
\end{figure}

In this work, we focus on demonstrating the feasibility of ultra low‑latency communication using an event‑camera‑based optical system. The proposed system employs a joint spatiotemporal modulation scheme. For clarity, we illustrate the modulation strategy using the $2 \times 4$ LED panel adopted in our experimental evaluation, as shown in Fig.~\ref{fig:testbed}. Although this prototype uses a limited number of LEDs, the achievable throughput scales linearly with the panel size. Larger LED arrays, such as a 1024‑LED panel, are planned in future work, along with more advanced modulation schemes to fully exploit the capabilities of higher‑resolution transmitters.
For the $2 \times 4$ configuration, each modulation symbol represents one byte (8 bits) and is encoded as the instantaneous state of the LED panel within a discrete time slot of duration $T = 1/f_{\text{bit}}$. Spatial encoding follows a row‑major indexing convention. The top‑left LED is assigned index~0 and corresponds to the most significant bit of each byte. Indices then increase from left to right and top to bottom, with the bottom‑right LED assigned index~7, representing the least significant bit. Within each symbol period, a binary value of~1 indicates that the corresponding LED is driven at its maximum light intensity (ON), while a value of~0 denotes zero intensity (OFF). Based on this design, we implement a robust spatial encoding and decoding framework that enables reliable symbol recovery under high‑speed, ultra‑low‑latency operation.

\subsection{Panel Localization and Device Association}

\begin{figure}[b]
    \centering
    \includegraphics[width=0.9\linewidth]{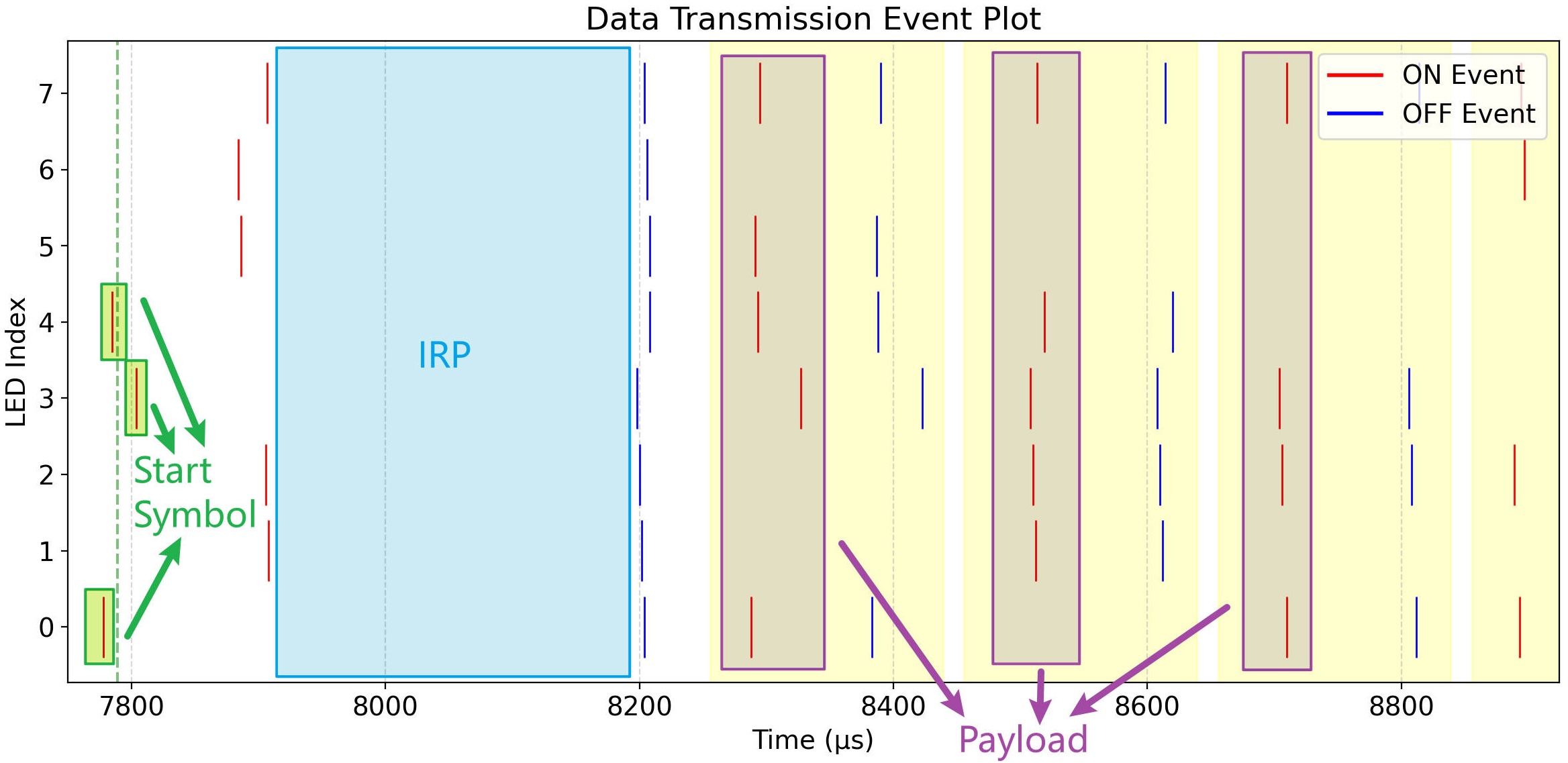}
    \vspace{-3mm}
    \caption{Demonstration on start symbol detection and payload symbols from the receiving events.}
    \label{fig:data_transmission}
\end{figure}

Although the proposed event‑camera‑based optical transmission system does not suffer from channel interference and therefore does not require contention‑resolution protocols such as CSMA, it must still address the challenges of device synchronization and reliable detection of data frame boundaries. This requirement becomes particularly critical in mobile scenarios, where the region of interest corresponding to each LED may shift dynamically within the sensor’s field of view. Inspired by conventional communication protocol design, we introduce a \emph{start symbol} that simultaneously serves as a panel localization marker and a data transmission indicator.
Specifically, the start symbol is encoded using the bit pattern \texttt{10011000} (hexadecimal \texttt{0x98}). In the $2 \times 4$ LED configuration, this pattern corresponds to the simultaneous activation of LEDs~0, 3, and~4, located at the top‑left, top‑right, and bottom‑left positions of the panel, respectively. This pattern exhibits strong spatial asymmetry and is analogous to the finder and alignment symbols used in QR codes. As a result, it provides a fixed spatial and temporal reference that allows the receiver to both synchronize to the beginning of a data frame and unambiguously determine the orientation and position of the LED panel within the sensor’s field of view. Notably, this combined localization and frame synchronization, or device association, process can be completed within the initial response period (IRP), typically within $1$~ms.
During idle periods when no data are transmitted, all LEDs remain in the OFF state. The receiver continuously monitors incoming events and searches for the predefined start symbol. Once detected, it computes the mean timestamp of the corresponding OFF‑to‑ON transition events and designates this value as the global frame start time, denoted by $t_{\text{start}}$. In addition, the detected spatial locations of the start‑symbol LEDs are used to infer the positions of the remaining LEDs on the panel, enabling consistent spatial decoding of all subsequent symbols within the data frame.

\subsection{Spatial Encoding}

\begin{algorithm}[b]
% \vspace{-3mm}
\caption{Data spatial encoding.}
\label{alg:spatial_encoding}
\begin{algorithmic}[1]

\State \textbf{Input:}
\State \quad $D$ : Data bytes $\{D_0, \dots, D_{n-1}\}$
\State \quad $T$ : Unit period
\State \quad $k$ : Number of waiting periods
\State \textbf{Output:}
\State \quad $S$ : Modulated Signal Bytes $\{S_0, \dots, S_{2n+k+1}\}$

\Statex \hrulefill

\Function{TransmitFrame}{$D, T, k$}
    \State \textbf{Start Symbol:} $S_{0}=$0x98
    \State \textbf{Waiting Periods:} $S_{1:k}=$0xFF
    \State \textbf{End of Waiting:} $S_{k+1}=$0x00
    \For{$i = 0$ \textbf{to} $n-1$}
        \State \textbf{Data byte:} $S_{2i+k+2}=D_i$
        \State \textbf{Slience Period:} $S_{2i+k+3}=$0x00
    \EndFor
    \State \Return $S$
\EndFunction

\end{algorithmic}
\end{algorithm}

The transmission procedure for a single data frame containing $n$ bytes of payload $D$ is summarized in Alg.~\ref{alg:spatial_encoding}. Each frame spans $(2n + k + 2)$ modulation periods, where $T=10^{6}/f_{\text{bit}}$~($\mu$s) denotes the duration of one period, and $S_i$ represents the LED panel state during $t \in [iT, (i+1)T)$. Transmission begins with a start symbol, followed by a sequence of control and data intervals.
Immediately after the start symbol, all remaining LEDs are turned ON for $k=\text{ceil}(250/T)$ consecutive periods to fully activate the corresponding pixel front ends. This wake‑up phase mitigates the initial response period (IRP) effect and reduces the likelihood of missing early data transitions, except at sufficiently low modulation frequencies (e.g., below $1{,}000$~bps), where IRP is negligible.
To maximize decoding robustness, each data byte is separated by a silence period during which all LEDs are OFF. Under this design, a binary value of~1 is always conveyed by an explicit OFF‑to‑ON transition, while a value of~0 produces no transition. This design allows us to further extend the tolerance window described in Section~\ref{sec:torlerance_window} to $[m \cdot T - T/2,, m \cdot T + 3T/2)$ to avoid possible cumulative readout delay. This transition‑based encoding aligns naturally with the operating principle of event cameras, which sense only intensity changes.

Although this structure introduces additional temporal redundancy, it prevents error propagation. Encoding bit~1 as a persistent ON state would require the receiver to track cumulative LED states, making the system vulnerable to missed or spurious events caused by noise or refractory effects. In contrast, alternating between data symbols and all‑OFF periods ensures that each byte is decoded independently, confining any error to a single period and mitigating inter‑symbol interference introduced by trailing effects.
Based on this framing, the total allocated duration of a frame is $(2n + k + 2)T$, which is the worst-case transmission latency. Under ideal conditions, the final effective data symbol occurs at $(2n + k)T$ after the start symbol. This property allows the receiver to finalize decoding immediately upon detecting the last data transition, thereby reducing end‑to‑end latency while maintaining robustness to sensor‑induced timing uncertainty.

\subsection{Decoding with Delay Compensation}

To mitigate readout bus congestion, we implement a delay compensation strategy that jointly leverages transmitter‑side encoding and receiver‑side decoding. This approach explicitly controls the timing of LED state transitions to regulate the temporal density of events arriving at the sensor.
As described in the IRP mitigation strategy, we introduce a deterministic, LED‑dependent delay $\Delta_k$ during encoding. For each modulation period $T$, instead of triggering all LEDs simultaneously, the state transition time for the LED with index $k$ is defined as
\begin{equation}
t_{k,\text{trigger}} = t_{\text{start}} + m \cdot T + k \cdot \Delta_{\text{base}},
\end{equation}
where $m$ denotes the modulation period index and $\Delta_{\text{base}}$ is a predefined base delay, for example $2,\mu\text{s}$. This intentional staggering distributes event generation across time, thereby reducing instantaneous contention on the sensor’s readout bus. By preventing a large burst of simultaneous events, the strategy alleviates readout saturation and reduces pixel‑dependent latency variation.

To recover the original bit pattern at the receiver, the applied temporal offsets must be compensated during decoding. For each event captured within the region of interest corresponding to LED $k$, the receiver computes a delay‑corrected timestamp
\begin{equation}
t_{\text{effective}} = t_{\text{measured}} - k \cdot \Delta_{\text{base}}.
\end{equation}
By subtracting the known deterministic delay, events originating from different LEDs are realigned to a common temporal reference. This compensation enables the decoder to apply a unified sampling window relative to the frame start time $t_{\text{start}}$ for all LEDs, independent of their spatial position or triggering order. As a result, the decoding process remains consistent with the original spatiotemporal encoding scheme while benefiting from reduced readout congestion and improved timestamp uniformity.

% To recover the original bit pattern on decoder side, the receiver must reverse this temporal shift. For every event captured within the ROI of LED $k$, we compute a compensated timestamp $t_{effective}$:
% \begin{equation}
%     t_{effective} = t_{measured} - k \cdot \Delta_{base}.
% \end{equation}
% By subtracting the known delay, the receiver aligns the events from all LEDs to a common time frame. This realignment allows the decoding logic to use a unified sampling window relative to the start timestamp of a frame $t_{start}$ for all LEDs.

\section{Evaluation}\label{sec:evalution}

\subsection{System Setup and Parameter Optimization}\label{sec:setup}

As shown in Fig.~\ref{fig:testbed}, the experimental platform consists of an optical transmitter and an event‑camera‑based receiver. The transmitter is implemented using a $2 \times 4$ LED panel driven by an FPGA development board, which provides nanosecond‑level timing precision for accurate control of modulation patterns and transmission frequencies. Leveraging this capability, the FPGA encodes data into spatiotemporal LED activation sequences and transmits them optically to the receiver.
The receiver is built around a LUCID Triton2 EVS camera. Although the lens supports variable zoom, the focal length is fixed across all experiments to ensure consistency. Sensor configuration and data acquisition are handled through the Arena SDK, which enables fine‑grained parameter control and high‑throughput event capture. To minimize latency artifacts in this prototype, events are buffered during acquisition and decoded offline into event tuples $(x, y, t, p)$, where $(x, y)$ denotes pixel location, $t$ the event timestamp, and $p$ the polarity. This offline pipeline ensures accurate and reproducible analysis of temporal and spatial event characteristics.

Reliable detection of high‑frequency LED modulation requires careful calibration of sensor parameters:
\begin{itemize}
    \item \textbf{Contrast thresholds:} Set the minimum intensity change for ON and OFF events; lower values improve sensitivity but increase noise.
    \item \textbf{Refractory period:} Controls effective pixel reset time via bias current; higher values enable faster event generation but do not affect the IRP (Section~\ref{sec:IRP}).
    \item \textbf{Threshold reference:} Determines front‑end bandwidth; higher values reduce latency and timestamp jitter, which is critical for high‑frequency modulation.
    \item \textbf{Event rate control (ERC):} Limits event throughput; disabled in all experiments to avoid event loss.
    \item \textbf{Region of interest (ROI):} Since only a single hardware ROI is supported, software filtering is applied to retain events from a $9 \times 9$ pixel area centered on each LED.
\end{itemize}
Due to page constraints and environment‑specific lighting conditions, exhaustive calibration procedures are omitted.

All subsequent evaluations are based on the following configurations. \textbf{I1:} A single LED is modulated using ON–OFF keying at bit rates from 1{,}000 to 20{,}000~bps. For each ON$\rightarrow$OFF and OFF$\rightarrow$ON transition, timestamps of all triggered events within the LED ROI are recorded and analyzed relative to the first event. Experiments are conducted at \emph{near} (0.3~m) and \emph{far} (1.1~m) distances to assess the impact of event density on temporal behavior.
\textbf{I2:} This configuration extends \textbf{I1} to eight LEDs modulated synchronously using the same pattern, enabling evaluation of readout contention and multi‑LED effects.

\subsection{Evaluation Results on Single-LED Time Stamp Inconsistency and Mitigation}

\begin{figure}[t]
    \centering
     % \vspace{-3mm}
    \subfigure[Near case.]{
        \includegraphics[width=0.4\linewidth]{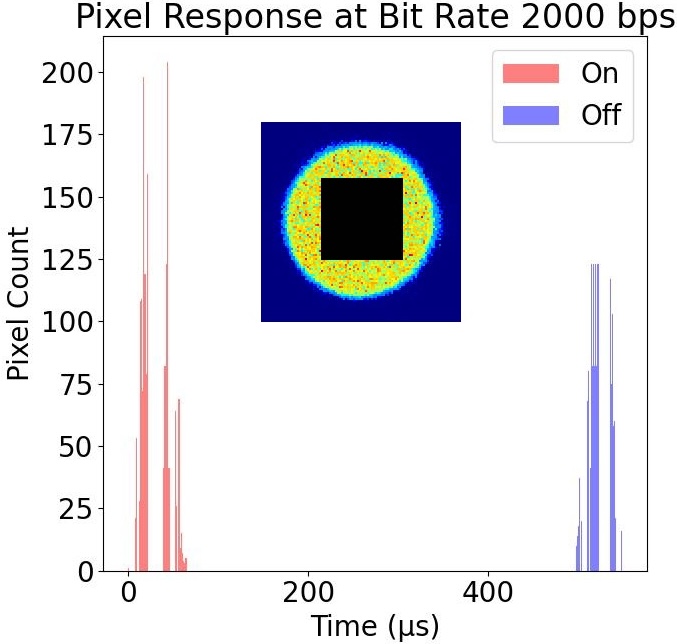}
        \label{fig:exp6_2_2_f2000_near_stacked}
    }%
     \hfill
    \subfigure[Far case.]{
        \includegraphics[width=0.4\linewidth]{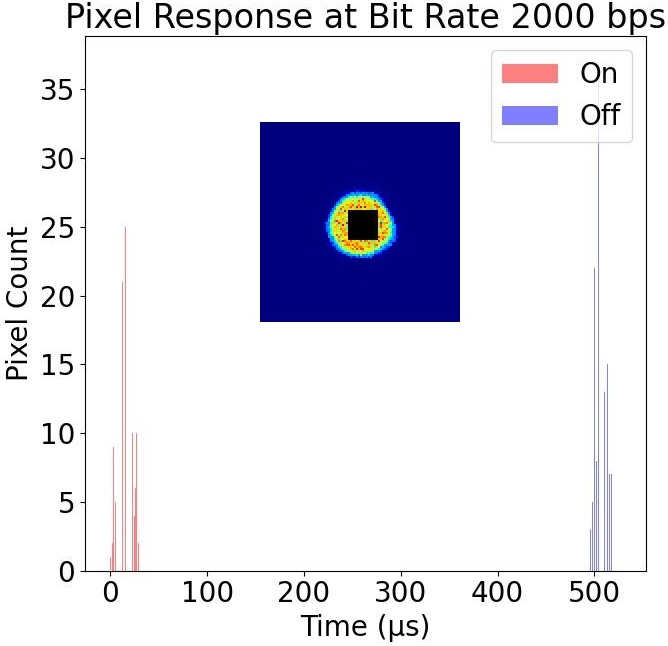}
        \label{fig:exp6_2_2_f2000_far_stacked}
    }
    \vspace{-5mm}
    \caption{ROI selection and corresponding pixel delay distributions at a modulation bit rate of 2000 bps. }
    \label{fig:exp6_2_2_event}
\end{figure}

\underline{Temporal consistency:}
We first evaluate temporal consistency with respect to \textbf{I1} by analyzing the LED pixel response delay, defined as the timestamp of the first event generated by each pixel following an LED state transition. As shown in Fig.~\ref{fig:exp6_2_2_event}, perspective effects cause the LED to illuminate a substantially larger sensor area at the near distance, corresponding to a $41 \times 41$ pixel ROI, compared to a $15 \times 15$ pixel ROI at the far distance.
For each transition, the earliest event within the ROI is used as a reference timestamp. The response delay is computed as the time difference between this reference and subsequent events in the same ROI, thereby isolating intra‑ROI delay variability while factoring out the absolute latency from the physical LED transition. This analysis is performed for both ON and OFF transitions at near and far distances to assess the impact of concurrent pixel activation on response latency.
As summarized in Table~\ref{tab:pixel_delay_near_far}, the near‑distance configuration consistently exhibits larger response delays than the far‑distance case. This behavior is attributed to readout bandwidth limitations: closer LEDs activate more pixels simultaneously, increasing contention on the sensor readout bus and inducing greater temporal dispersion. We further observe that minimizing the refractory period significantly reduces these delays. Together, these results underscore the importance of controlling event volume and refractory settings to achieve stable, high‑resolution temporal measurements under high‑frequency optical stimulation.

% \begin{figure}[ht!]
%     \centering
%     \subfigure[ON events.]{
%         \includegraphics[width=0.48\linewidth]{figures/exp6_2_2_response_delay_on.jpg}
%         \label{fig:exp6_2_2_response_delay_on}
%     }% 
%     \hfill
%     \subfigure[OFF events.]{
%         \includegraphics[width=0.48\linewidth]{figures/exp6_2_2_response_delay_off.jpg}
%         \label{fig:exp6_2_2_response_delay_off}
%     }
%     \caption{Pixel response delay ($\mu s$) vs. transmit bit rate.}
%     \label{fig:exp6_2_2_response_delay}
% \end{figure}

\begin{table}[ht!]
\centering
% \scriptsize
\small
\caption{Pixel response delay ($\mu s$) comparison.}
\label{tab:pixel_delay_near_far}
\setlength{\tabcolsep}{3pt}
\begin{tabular}{c|cc|cc|cc|cc}
\hline
  & \multicolumn{2}{c|}{\textbf{ON at 0.3 m}} & \multicolumn{2}{c|}{\textbf{OFF at 0.3 m}} & \multicolumn{2}{c|}{\textbf{ON at 1.1 m}} & \multicolumn{2}{c}{\textbf{OFF at 1.1 m}} \\
\cline{2-9}
\textbf{Bps} 
& \textbf{Mean} & \textbf{Std} & \textbf{Mean} & \textbf{Std}
& \textbf{Mean} & \textbf{Std} & \textbf{Mean} & \textbf{Std} \\
\hline
1000 & 23.8 & 12.0 & 43.7 & 14.4 & 15.1 & 6.8 & 14.8 & 6.9 \\
2000 & 21.5 & 11.1 & 26.5 & 14.6 & 15.1 & 6.7 & 14.5 & 6.7 \\
3000 & 24.7 & 11.3 & 23.1 & 10.0 & 14.4 & 6.8 & 12.2 & 6.4 \\
4000 & 20.9 & 10.3 & 21.4 & 10.4 & 15.1 & 6.8 & 11.9 & 7.6 \\
5000 & 21.0 & 10.2 & 22.4 & 10.5 & 13.8 & 7.1 & 12.7 & 6.2 \\
8000 & 25.9 & 10.8 & 23.3 & 11.1 & 15.7 & 7.6 & 12.2 & 6.9 \\
12000 & 19.2 & 10.9 & 24.0 & 10.3 & 17.3 & 7.5 & 17.9 & 8.2 \\
16000 & 19.5 & 8.9 & 22.0 & 10.3 & 16.6 & 8.2 & 18.9 & 9.2 \\
20000 & 15.7 & 8.6 & 13.2 & 7.2 & 19.7 & 7.9 & 20.6 & 8.6 \\
\hline
\end{tabular}
\end{table}

\begin{figure}[ht!]
    \centering
    \vspace{-3mm}
    \subfigure[Near case.]{
        \includegraphics[width=0.48\linewidth]{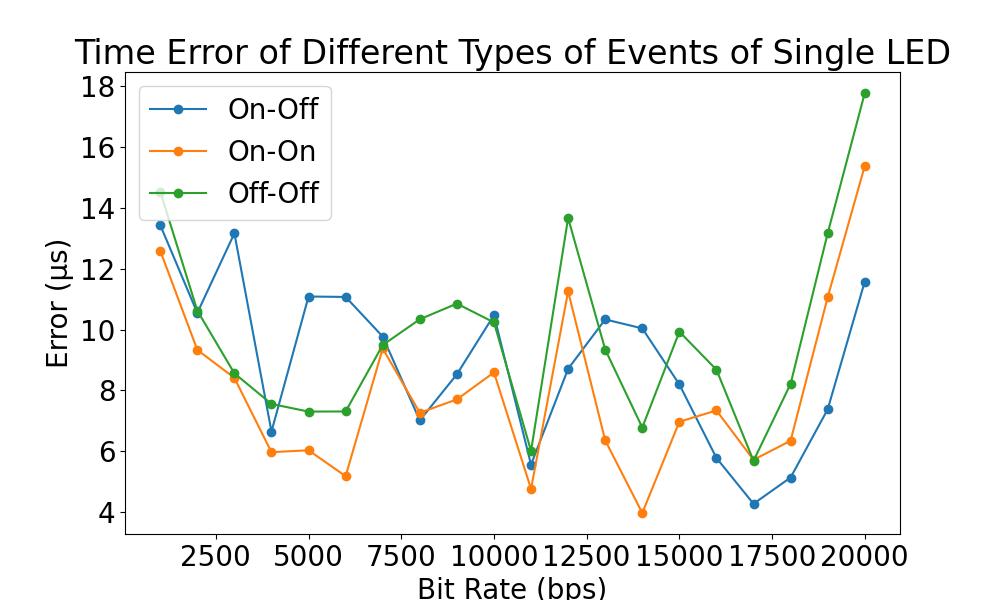}
        \label{fig:exp6_2_1_near_time_error_n2}
    }%
    \hfill
    \subfigure[Far case.]{
        \includegraphics[width=0.48\linewidth]{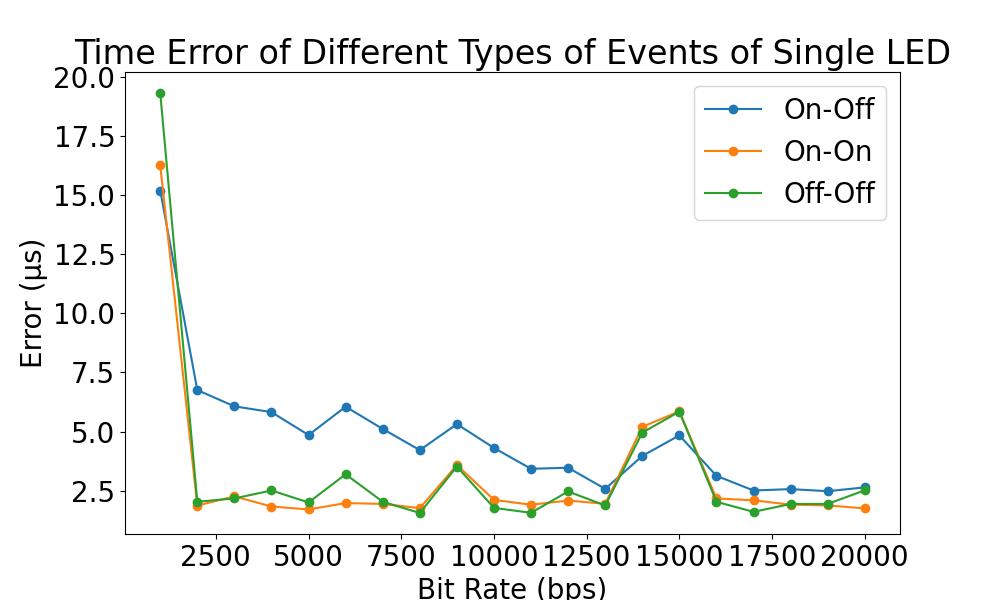}
        \label{fig:exp6_2_1_far_time_error_n2}
    }
    \vspace{-3mm}
    \caption{Measurement error ($\mu s$) vs. transmit bit rate with ON-ON, ON-OFF, OFF-OFF event pairs.}
    \label{fig:exp6_2_1_time_error}
\end{figure}

\underline{Evaluation on ON and OFF events:}
We next compare the sensor’s sensitivity and timing accuracy for ON and OFF events under \textbf{I1}. Events within each LED ROI are aggregated using the S‑2 method, which provides the best empirical performance (see Section~\ref{sec:multi-evalution}). We measure temporal intervals between successive events of the same and different polarities, including ON‑to‑ON, OFF‑to‑OFF, and cross‑polarity ON‑to‑OFF (and OFF‑to‑ON) pairs. These measured intervals are compared against the expected timing derived from the modulation bit rate, and the deviation is used to quantify response asymmetry to rising and falling illumination at near and far distances.
As shown in Fig.~\ref{fig:exp6_2_1_time_error}, ON‑to‑ON event pairs consistently exhibit lower timing error than OFF‑to‑OFF pairs. This result indicates higher sensitivity and more stable temporal response for ON transitions. In contrast, ON‑to‑OFF intervals show no consistent trend, as their accuracy depends on the accumulated timing errors of both the preceding ON and OFF events.

\begin{figure}[ht!]
    \centering
    \vspace{-2mm}
    \subfigure[Near case.]{
        \includegraphics[width=0.48\linewidth]{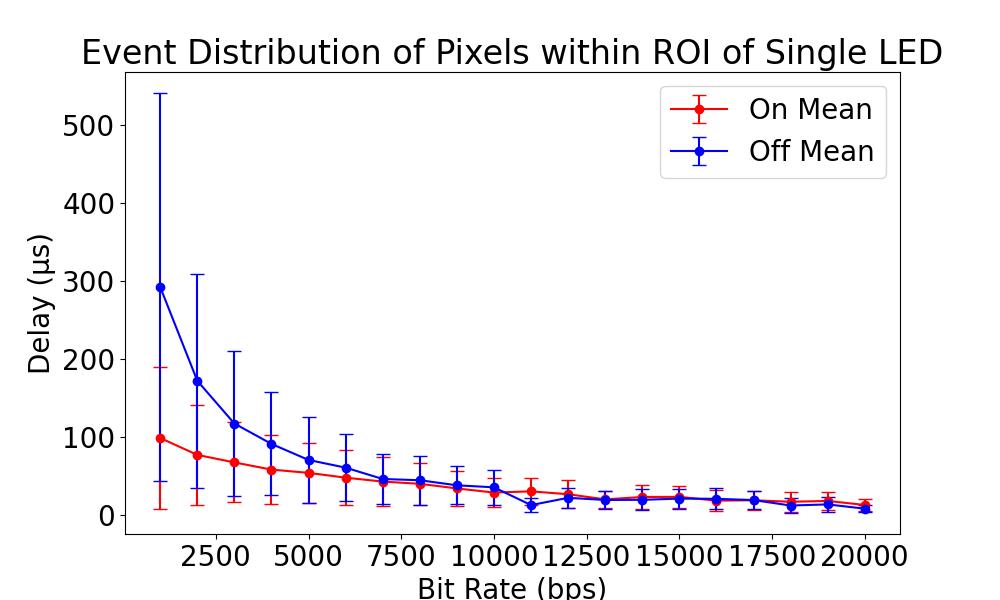}
        \label{fig:exp6_2_1_near_trailing_effect}
    }%
    \hfill
    \subfigure[Far case.]{
        \includegraphics[width=0.48\linewidth]{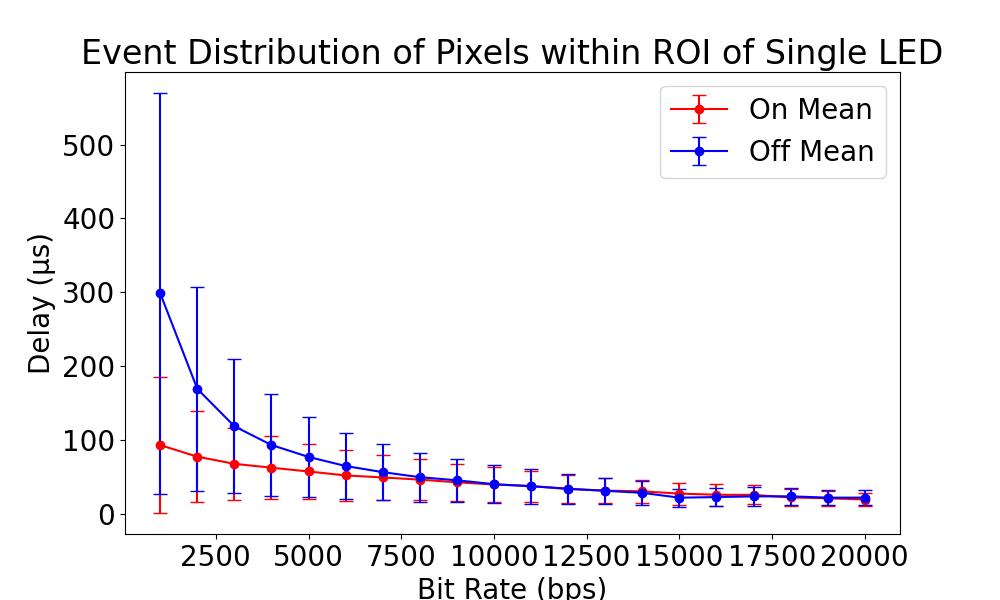}
        \label{fig:exp6_2_1_far_trailing_effect}
    }
    \vspace{-3mm}
    \caption{Event distribution timestamp ($\mu s$) of ON and OFF events vs. transmit bit rate.}
    \label{fig:exp6_2_1_trailing_effect}
    \vspace{-4mm}
\end{figure}

\underline{Trailing effect:} We further investigate the trailing effect under \textbf{I1}. As illustrated in Fig.~\ref{fig:exp6_2_1_trailing_effect}, trailing behavior is observed for both ON and OFF events. OFF transitions generally produce longer trailing windows at lower bit rates. As the bit rate increases, the observed trailing duration decreases, primarily because subsequent transitions occur sooner and truncate the ongoing response. The close agreement between near and far measurements suggests that trailing is dominated by pixel‑level analog front‑end recovery dynamics rather than readout‑level congestion effects.

% \subsection{Inevitable Refractory Period (IRP) at Different Bit Rates}

\underline{IRP:} To evaluate the IRP under standard modulation, we performed an experiment based on \textbf{I1} at 1.1 m distance across a range of bit rates from 1 kbps to 20 kbps. We intentionally trigger IRP by adding a ``silent'' period where the LED remained OFF for a long duration (e.g., 0.5 s) before operation. We recorded the first ON event generated after the long pause and measured the time interval between this initial ON event and the subsequent first OFF event. This interval represents the IRP of the pixels within the region responsive to the LED immediately after waking from an idle state, helping us quantify the transmission-incapable duration introduced by IRP across different bit rates. Ideally, assuming no IRP, the interval shall be  $T = 10^6/f_{\text{bit}}~(\mu s)$. We define the IRP temporal latency error $E_{IRP}(f)$ as:
\begin{equation}\label{eq:irp_error}
    E_{\text{IRP}}(f) = \Delta t_{\text{measured}} - T.
\end{equation}
As shown in Table~\ref{tab:exp6_5_1_irp}, at bit rates lower than or equal to 4,000 bps, the measured ON-to-OFF intervals are close to the theoretical values, indicating that the signal period is long enough such that it is almost unaffected by the photosensor's bandwidth limit. However, as the bit rate increases beyond 4,000 bps, the measured interval deviates sharply from the theoretical curve and gradually increases. This behavior shows that at higher bit rates, the photosensor's built-in low-pass filter attenuates more on the sharp changes of signal. Consequently, the differencing circuit requires a longer duration to accumulate enough change to setup a reference point, causing a longer IRP as the bit rate approaches the analog front-end's cutoff limit.
\begin{table}[ht!]
    \centering
    \small
    \caption{IRP and error for different transmission rates (1.1m).}
    \label{tab:exp6_5_1_irp}
    \begin{tabular}{c|c|c|c|c}
\hline
\textbf{Bps} & \textbf{Mean diff} & \textbf{Std diff} & \textbf{Expected} & \textbf{Error ($\mu s$)} \\
\hline
1000 & 1002.78 & 5.29 & 1000 & 2.78 \\
\hline
2000 & 497.40 & 44.25 & 500 & 2.60 \\
% \hline
% 3000 & 336.50 & 4.26 & 333 & 3.50 \\
\hline
4000 & 264.14 & 6.39 & 250 & 14.14 \\
% \hline
% 5000 & 249.79 & 7.57 & 200 & 49.79 \\
\hline
6000 & 267.63 & 10.64 & 166 & 101.63 \\
% \hline
% 7000 & 427.53 & 40.64 & 142 & 285.53 \\
\hline
8000 & 381.53 & 4.43 & 125 & 256.53 \\
\hline
% 9000 & 351.73 & 4.65 & 111 & 240.73 \\
% \hline
10000 & 331.10 & 6.12 & 100 & 231.10 \\
\hline
% 11000 & 325.38 & 14.33 & 90 & 235.38 \\
% \hline
12000 & 376.39 & 52.80 & 83 & 293.39 \\
\hline
% 13000 & 397.78 & 5.24 & 76 & 321.78 \\
% \hline
14000 & 375.98 & 4.70 & 71 & 304.98 \\
\hline
% 15000 & 361.12 & 5.44 & 66 & 295.12 \\
% \hline
16000 & 351.82 & 11.18 & 62 & 289.82 \\
\hline
% 17000 & 401.08 & 38.86 & 58 & 343.08 \\
% \hline
18000 & 400.70 & 34.73 & 55 & 345.70 \\
\hline
% 19000 & 388.73 & 4.78 & 52 & 336.73 \\
% \hline
20000 & 374.27 & 4.45 & 50 & 324.27 \\
\hline
\end{tabular}
\end{table}
We further modify the modulation pattern to investigate the exact duration of IRP. After a long \emph{silent} pause (0.5 s), the single LED iss turned ON and held in the ON state for a variable duration before being turned OFF. As shown in Table~\ref{tab:exp6_5_2_irp}, the interval error drops to a value close to zero when the hold time exceeds 250 $\mu s$. This indicates that the IRP for the sensor is approximately 250 $\mu s$. Once the pixel has been active for this duration, the internal differencing circuit and comparator reach a stable state, allowing the sensor to respond accurately to subsequent transitions.

% \begin{figure}[ht!]
%     \centering
%     \includegraphics[width=0.8\linewidth]{figures/exp6_5_2_irp.jpg}
%     \caption{IRP and mean error for different hold time.}
%     \label{fig:exp6_5_2_irp}
% \end{figure}

\begin{table}[ht!]
    \centering
    \small
    \caption{Results on IRP evaluation for different hold time.}
    \label{tab:exp6_5_2_irp}
    \begin{tabular}{c|c|c|c|c}
\hline
\textbf{Hold ($\mu s$)} & \textbf{Mean diff} & \textbf{Std diff} & \textbf{Expected} & \textbf{Error ($\mu s$)} \\
\hline
100 & 584.45 & 52.23 & 100 & 484.45 \\
\hline
150 & 269.78 & 8.05  & 150 & 119.78 \\
\hline
200 & 243.43 & 3.92  & 200 & 43.43  \\
\hline
250 & 255.25 & 3.94  & 250 & 5.25   \\
\hline
300 & 303.91 & 3.94  & 300 & 3.91   \\
\hline
350 & 350.08 & 1.86  & 350 & 0.08   \\
\hline
400 & 402.60 & 4.18  & 400 & 2.60   \\
\hline
450 & 450.45 & 2.50  & 450 & 0.45   \\
\hline
500 & 503.67 & 2.72  & 500 & 3.67   \\
\hline
550 & 551.00 & 3.87  & 550 & 1.00   \\
\hline
600 & 603.85 & 4.65  & 600 & 3.85   \\
\hline
\end{tabular}
\end{table}

\subsection{Evaluation Results on Multi-LED Time Stamp Inconsistency and Mitigation}\label{sec:multi-evalution}

We next evaluate the event aggregation strategy under the multi‑LED configuration \textbf{I2}. Although all LEDs are programmed to switch states synchronously, the event camera inherently records their responses asynchronously due to sequential readout and contention effects. To address this, events within each LED’s ROI are aggregated to infer the corresponding state transition using five aggregation methods: Best Pixel, S‑2, S‑5, S‑10, and Cluster‑based aggregation (see Section~\ref{sec:aggregation}). For each transition, the LED with the earliest aggregated response is selected as a temporal reference, and relative delays of the remaining LEDs are computed with respect to this reference. Experiments are conducted for both near and far distances.
As shown in Table~\ref{tab:aggregation_delay}, the S‑2 aggregation method achieves a favorable balance between timing accuracy and computational efficiency, making it well suited for real‑time processing. In addition, the far‑distance configuration consistently results in lower inter‑LED delays than the near‑distance case, as the reduced event density per ROI alleviates readout bus congestion. These results highlight the combined importance of physical setup and aggregation strategy in achieving reliable multi‑source temporal synchronization with event cameras.

% \begin{figure}[ht!]
%     \centering
%     \subfigure[At 0.3 m.]{
%         \includegraphics[width=0.48\linewidth]{figures/exp6_3_1_near.jpg}
%         \label{fig:exp6_3_1_near}
%     }%
%     \hfill
%     \subfigure[At 1.1 m.]{
%         \includegraphics[width=0.48\linewidth]{figures/exp6_3_1_far.jpg}
%         \label{fig:exp6_3_1_far}
%     }
%     \caption{Event density and mean delay ($\mu s$) results, and sensor readout plots.}
%     \label{fig:exp6_3_1}
% \end{figure}

\begin{table}[t]
\centering
\small
\setlength{\tabcolsep}{3pt}
\caption{Inter-LED mean delay ($\mu$s) for different decoding methods in near and far cases.}
\label{tab:aggregation_delay}
\begin{tabular}{c|ccccc|ccccc}
\hline
& \multicolumn{5}{c|}{\textbf{Near (0.3 m)}} & \multicolumn{5}{c}{\textbf{Far (1.1 m)}} \\
\cline{2-11}
\textbf{Bps}
& \textbf{BP} & \textbf{S-2} & \textbf{S-5} & \textbf{S-10} & \textbf{Clu}
& \textbf{BP} & \textbf{S-2} & \textbf{S-5} & \textbf{S-10} & \textbf{Clu} \\
\hline
1000 & 18.16 & 17.30 & 17.30 & 17.27 & 17.27 & 9.48 & 9.13 & 9.00 & 8.54 & 9.56 \\
2000 & 17.59 & 16.90 & 17.10 & 16.75 & 16.22 & 9.02 & 7.78 & 7.41 & 7.12 & 8.04 \\
% 3000 & 15.18 & 15.89 & 18.56 & 18.89 & 15.73 & 10.35 & 7.27 & 7.27 & 7.18 & 7.67 \\
4000 & 16.22 & 15.78 & 15.71 & 17.09 & 15.84 & 9.89 & 7.64 & 7.04 & 7.25 & 7.50 \\
% 5000 & 15.20 & 15.24 & 15.00 & 14.94 & 15.27 & 7.47 & 7.05 & 6.77 & 7.07 & 7.27 \\
6000 & 16.05 & 21.18 & 17.10 & 16.94 & 18.25 & 9.52 & 7.50 & 6.64 & 6.52 & 7.25 \\
% 7000 & 15.55 & 15.67 & 16.06 & 15.65 & 15.49 & 8.03 & 6.97 & 6.46 & 6.64 & 7.06 \\
8000 & 18.31 & 20.23 & 20.23 & 20.29 & 18.18 & 9.95 & 6.50 & 6.59 & 6.29 & 6.76 \\
% 9000 & 15.68 & 14.91 & 14.91 & 15.25 & 15.18 & 11.41 & 6.86 & 6.23 & 6.03 & 6.95 \\
10000 & 16.93 & 15.23 & 15.69 & 15.57 & 15.09 & 12.53 & 7.55 & 5.55 & 5.57 & 6.76 \\
% 11000 & 19.04 & 15.45 & 15.33 & 17.35 & 15.50 & 10.97 & 6.31 & 6.07 & 5.96 & 7.71 \\
12000 & 42.67 & 17.64 & 17.94 & 17.96 & 17.88 & 12.54 & 5.90 & 5.62 & 5.77 & 6.44 \\
% 13000 & 16.52 & 24.02 & 18.61 & 21.18 & 22.20 & 11.73 & 7.88 & 8.51 & 6.79 & 9.39 \\
14000 & N/A & 17.17 & 19.17 & 18.05 & 15.03 & 10.37 & 6.83 & 6.12 & 6.20 & 6.30 \\
% 15000 & 16.79 & 17.37 & 14.48 & 15.43 & 16.69 & 12.96 & 5.73 & 6.24 & 5.57 & 5.48 \\
16000 & N/A & 33.05 & 39.33 & 16.05 & 42.57 & 10.67 & 5.98 & 6.30 & 5.65 & 6.98 \\
% 17000 & N/A & 46.24 & 37.07 & 47.64 & 46.10 & 11.86 & 10.48 & 7.37 & 6.90 & 7.73 \\
18000 & N/A & 55.64 & 55.71 & 55.79 & 41.81 & 16.83 & 6.54 & 6.05 & 6.34 & 6.36 \\
% 19000 & 59.43 & 55.71 & 55.71 & 55.43 & 55.29 & 17.49 & 8.20 & 6.24 & 6.29 & 6.03 \\
20000 & N/A & 55.86 & 55.95 & 56.52 & 56.14 & 18.34 & 5.93 & 7.33 & 7.63 & 5.70 \\
\hline
\end{tabular}
\vspace{-3mm}
\end{table}

\begin{figure}[b]
    \centering
    \vspace{-6mm}
    \subfigure[2 LEDs.]{
        \includegraphics[width=0.45\linewidth]{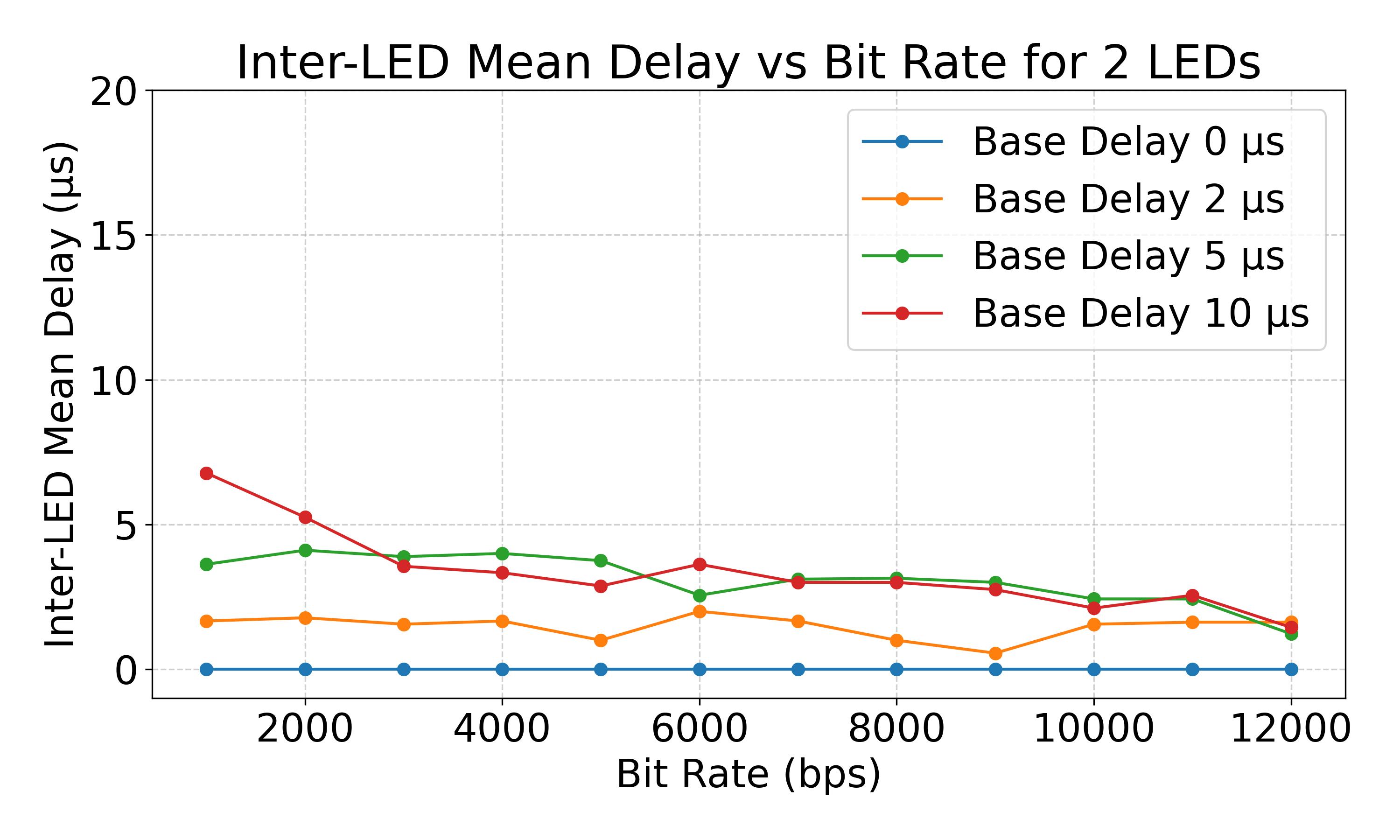}
        \label{fig:exp6_3_2_far_led2}
    }
    \hfill
    \subfigure[4 LEDs.]{
        \includegraphics[width=0.45\linewidth]{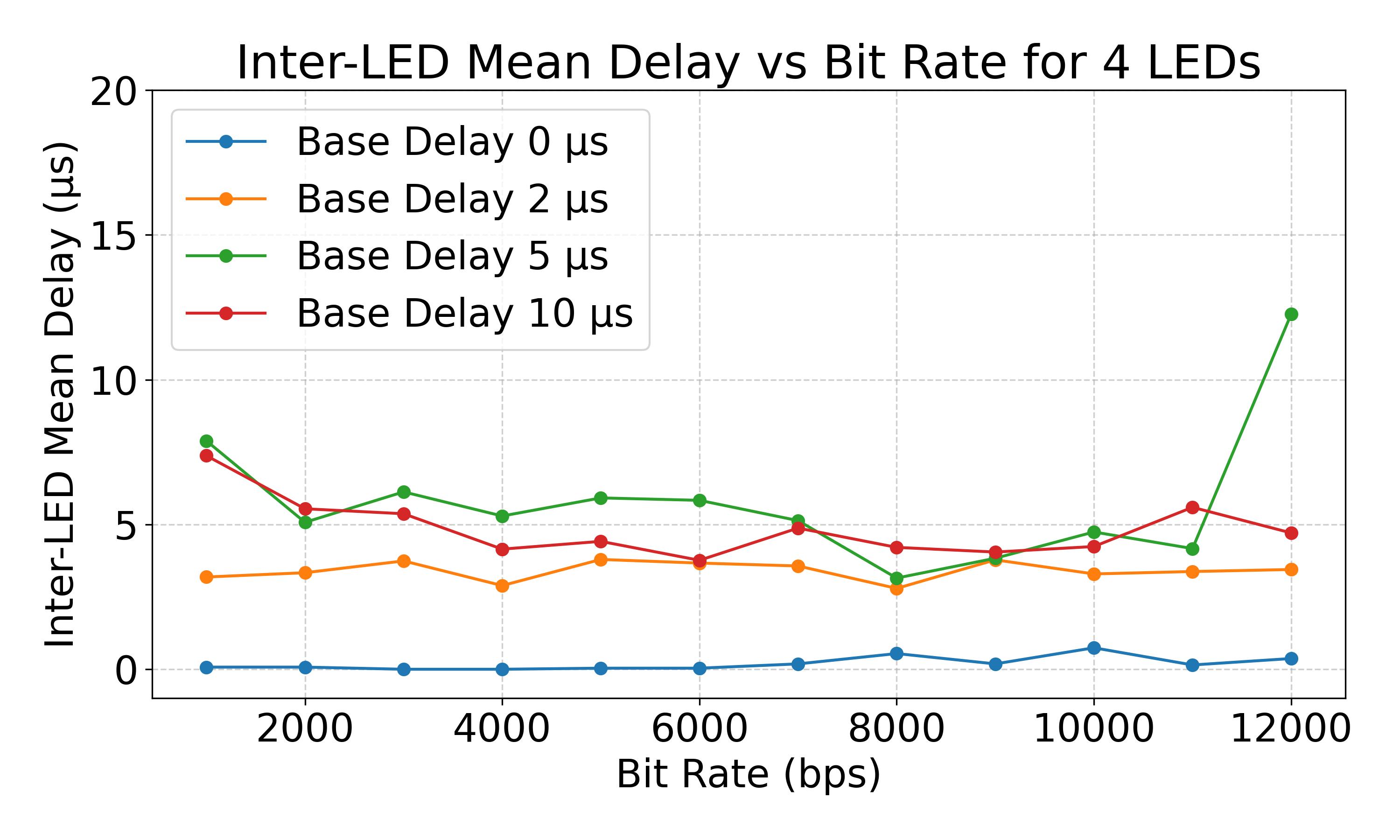}
        \label{fig:exp6_3_2_far_led4}
    }

 \vspace{-3mm}
    \subfigure[6 LEDs.]{
        \includegraphics[width=0.45\linewidth]{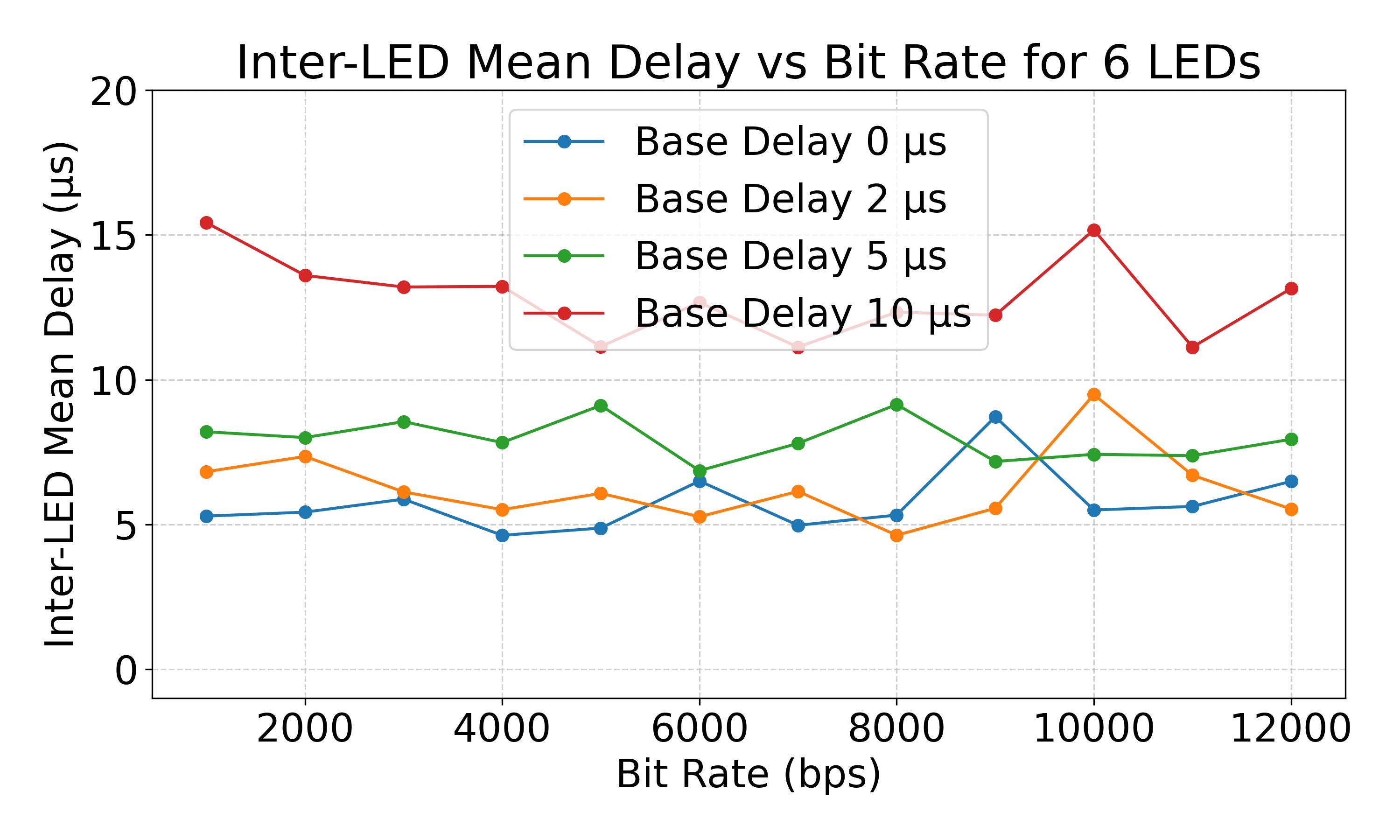}
        \label{fig:exp6_3_2_far_led6}
    }
    \hfill
    \subfigure[8 LEDs.]{
        \includegraphics[width=0.45\linewidth]{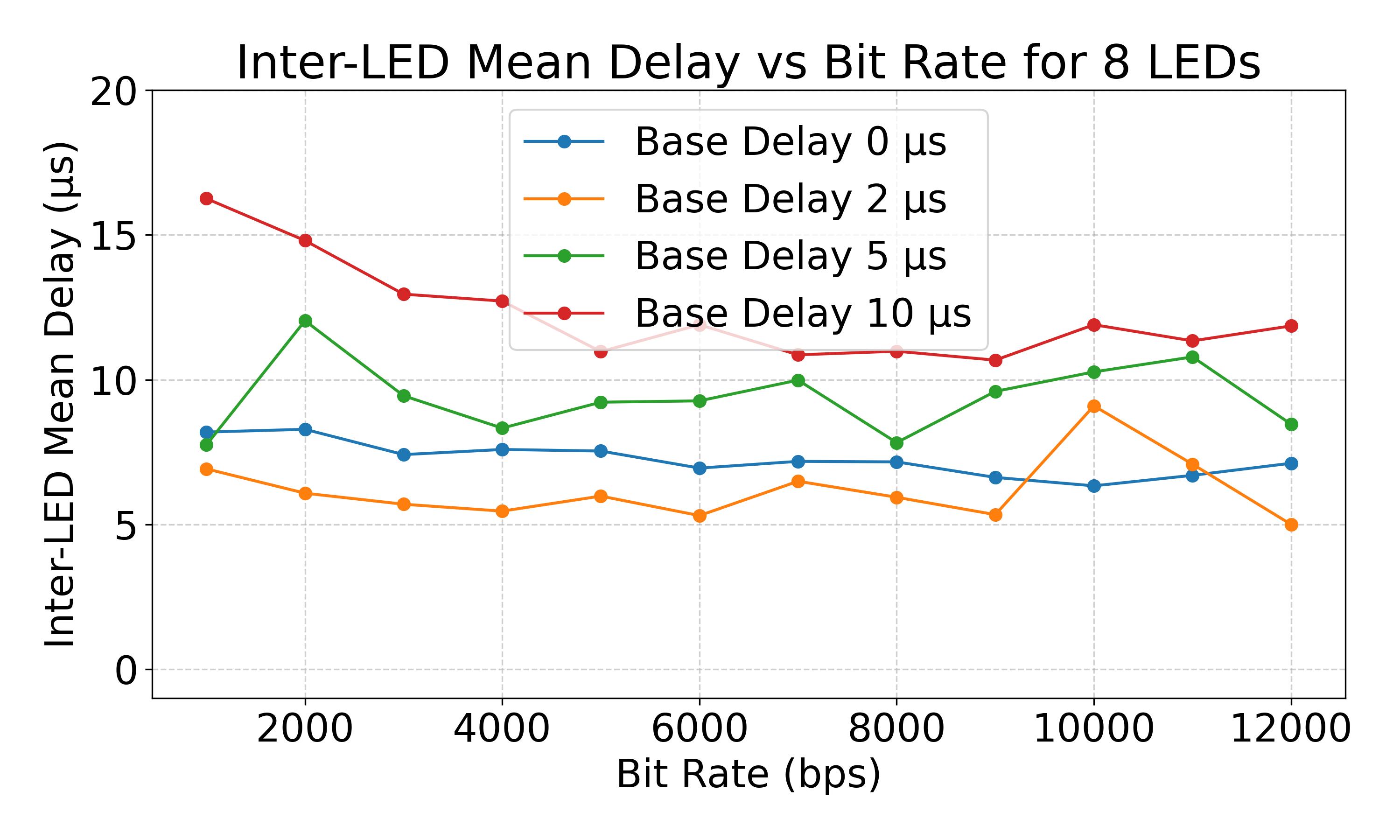}
        \label{fig:exp6_3_2_far_led8}
    }
    \vspace{-4mm}
    \caption{Mean delay ($\mu$s) comparisons for different delays with different active LEDs.}
    \label{fig:exp6_3_2_delay_across_number}
\end{figure}

We further investigate the impact of the number of active LEDs on inter‑LED readout delay. Extending \textbf{I2}, we vary the number of simultaneously active LEDs from 2 to 8 while keeping all other parameters fixed, and aggregate events using the S‑2 method. For each configuration, relative delays among LEDs are measured to capture the relationship between event volume and readout bus congestion. All experiments are conducted in the far‑distance setup to isolate the effect of LED count from spatial extent, and delay compensation is applied to improve alignment.
As shown in Fig.~\ref{fig:exp6_3_2_delay_across_number}, increasing the number of active LEDs results in a clear increase in mean readout delay. This trend reflects the finite readout bandwidth of the sensor: while additional LEDs enrich spatial information, they also generate proportionally more events. As the event rate increases, the sequential arbitration of the readout bus forces later‑processed regions to wait longer, inflating recorded timestamps and increasing temporal dispersion across LEDs.

\begin{figure}[t]
    \centering
     % \vspace{-3mm}
    \subfigure[LEDs 0 \& 1.]{
        \includegraphics[width=0.45\linewidth]{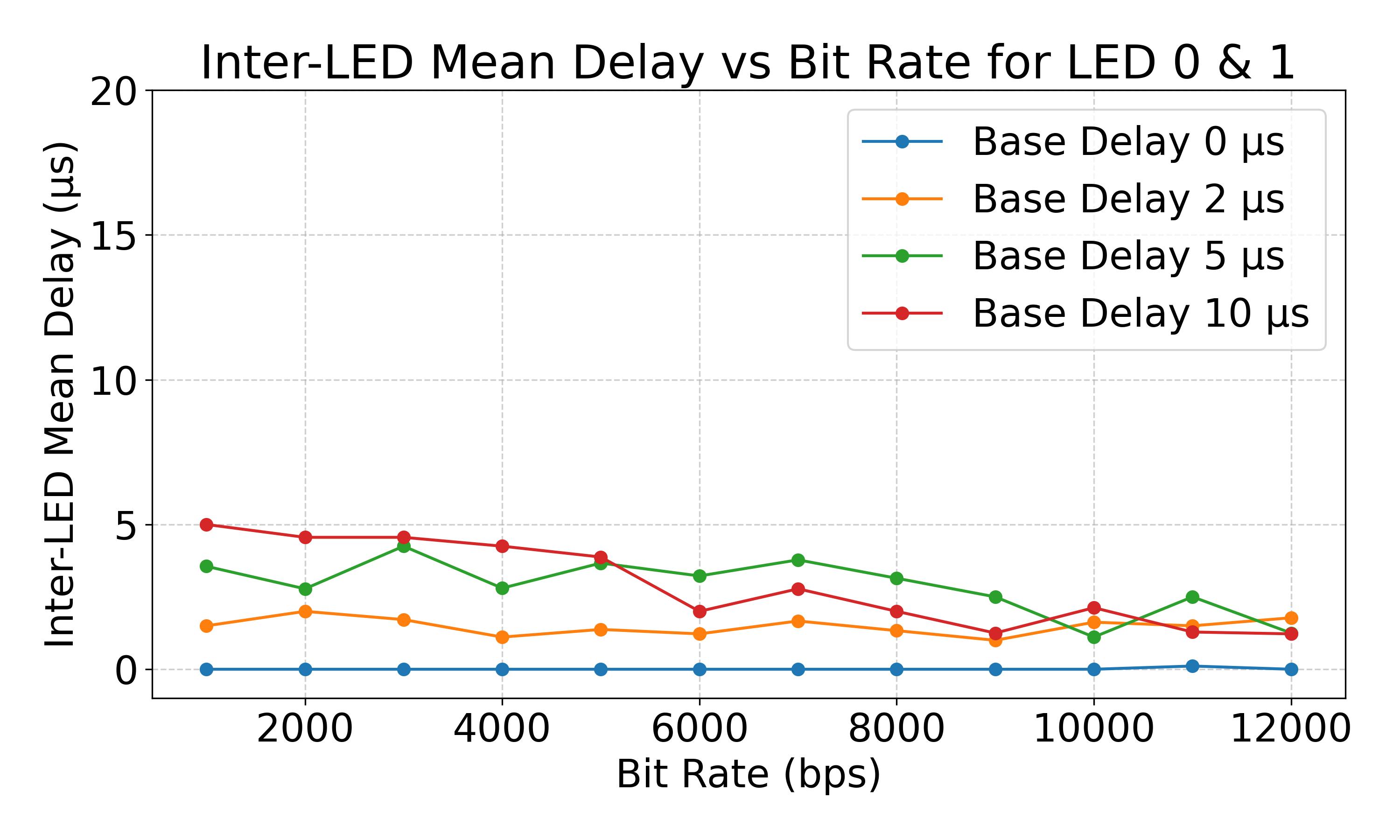}
        \label{fig:exp6_3_3_led_far_0_1}
    }
    \hfill
    \subfigure[LEDs 0 \& 3.]{
        \includegraphics[width=0.45\linewidth]{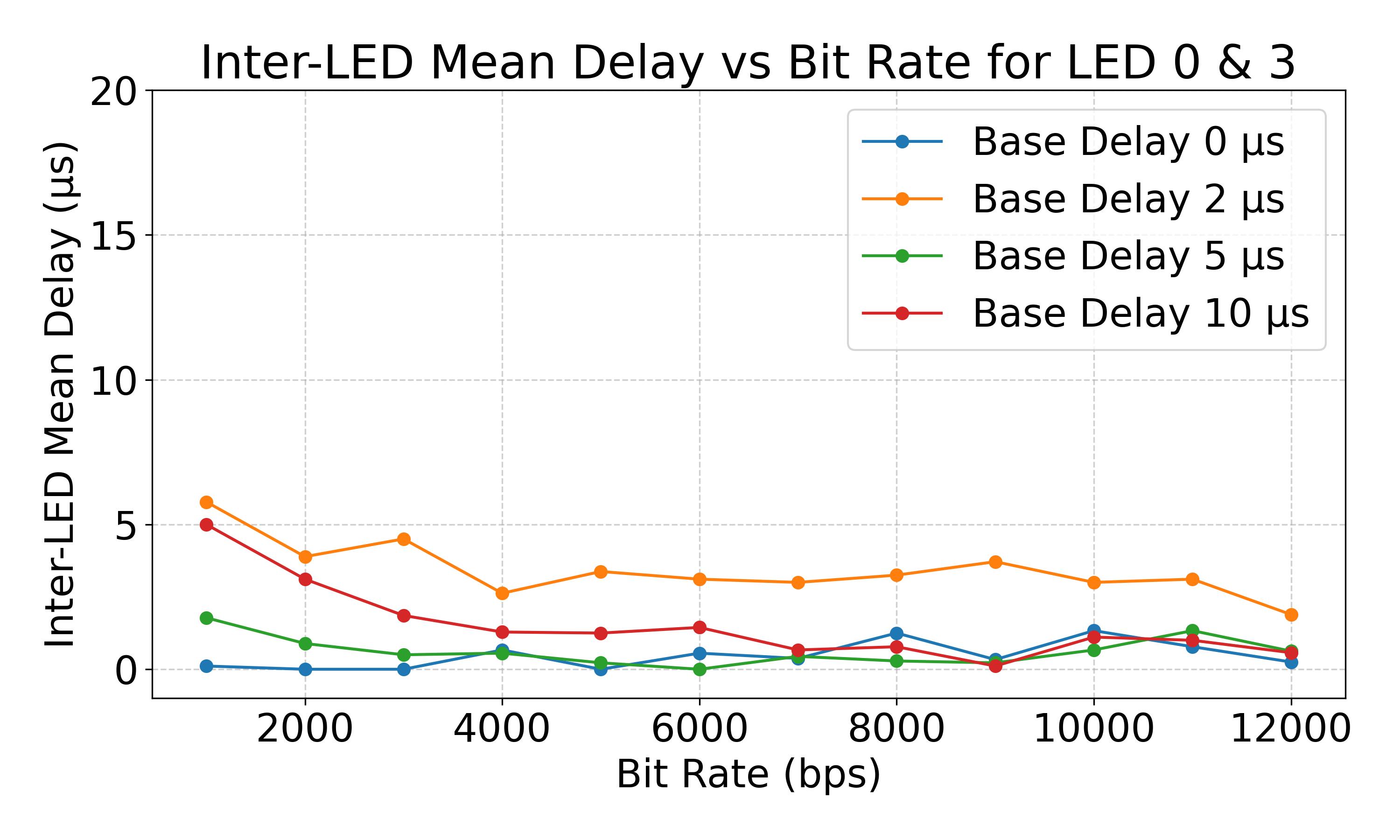}
        \label{fig:exp6_3_3_led_far_0_3}
    }

 \vspace{-3mm}
    \subfigure[LEDs 0 \& 4.]{
        \includegraphics[width=0.45\linewidth]{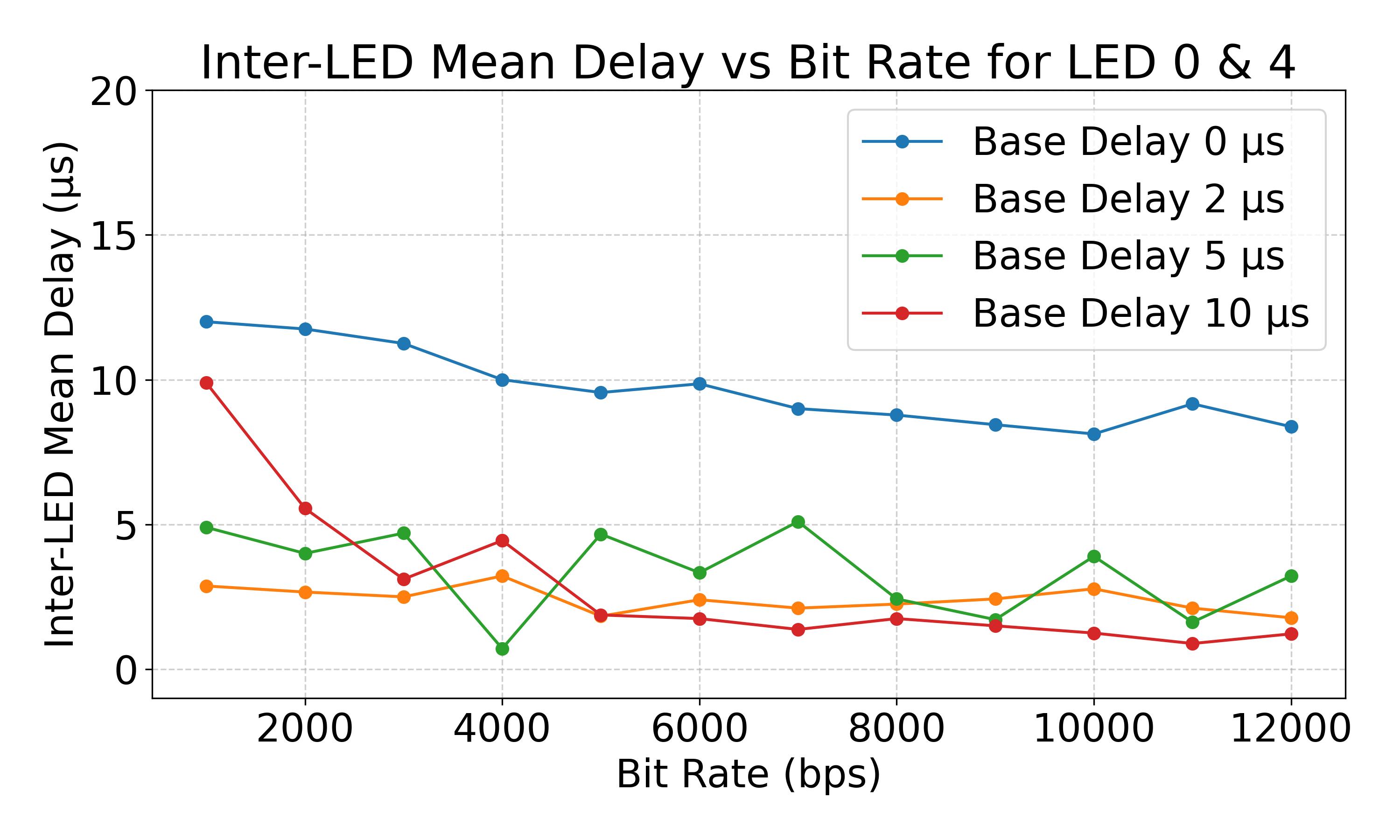}
        \label{fig:exp6_3_3_led_far_0_4}
    }
    \hfill
    \subfigure[LEDs 0 \& 7.]{
        \includegraphics[width=0.45\linewidth]{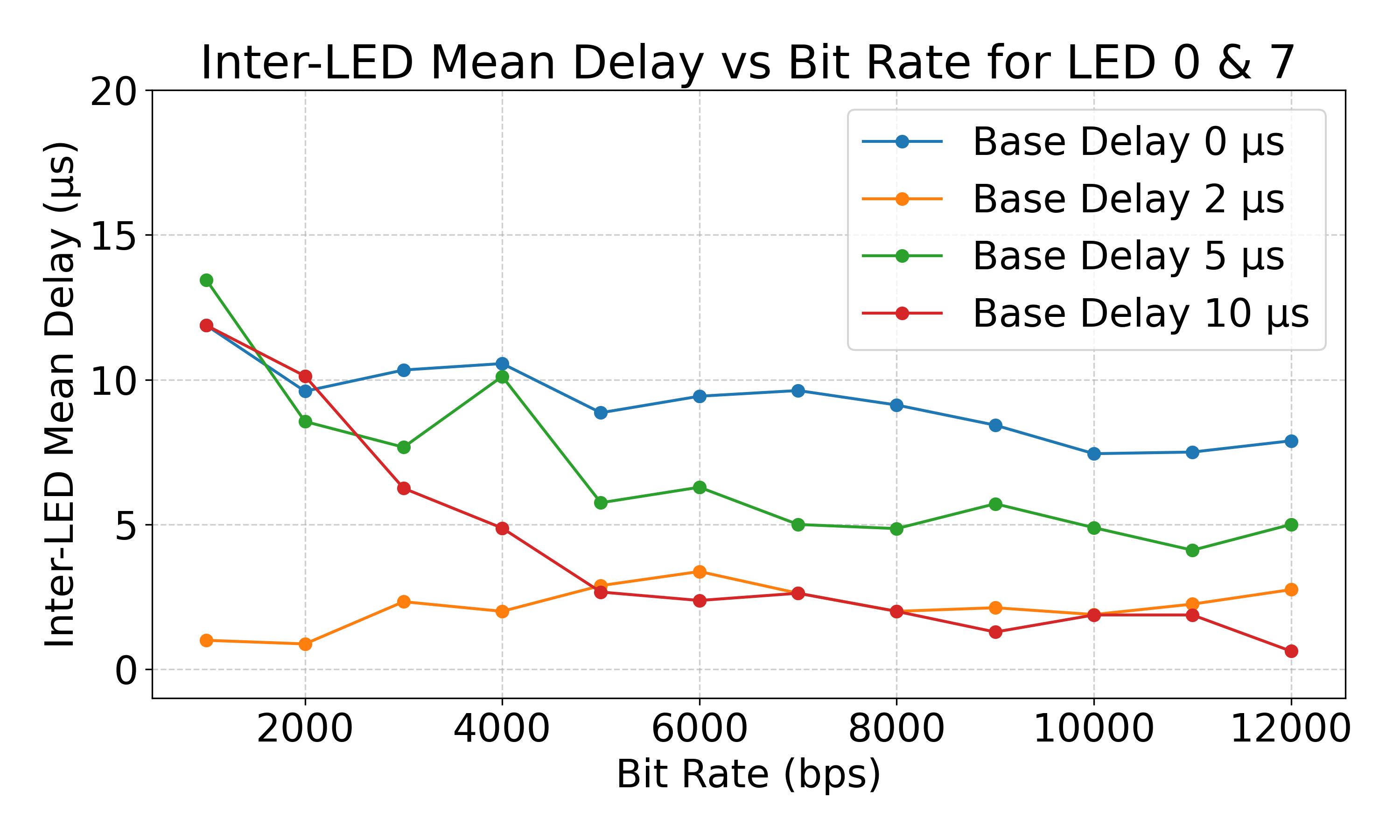}
        \label{fig:exp6_3_3_led_far_0_7}
    }
    \vspace{-4mm}
    \caption{Mean delay ($\mu$s) comparisons for different delays at various LED pairs.}
    \label{fig:exp6_3_3_delay_across_pair}
    \vspace{-4mm}
\end{figure}

Building on the observation that readout delays depend on spatial location, we further examine how the relative positions of active regions on the sensor affect temporal consistency. In most cases, the sensor reads all pixels in a row at a time, thereby the horizontal LEDs will have a lower relative delay than vertical ones. Using configuration \textbf{I2} with S‑2 aggregation, we focus on scenarios with exactly two active LEDs and evaluate multiple LED pairings, specifically LED 0 paired with LEDs 1, 3, 4, and 7. These combinations represent increasing spatial separation on the sensor.
For each pairing, we measure the relative delay between the two LEDs across a range of modulation bit rates and in far-distance setup. As shown in Fig.~\ref{fig:exp6_3_3_delay_across_pair}, inter‑LED readout delay exhibits a strong correlation with spatial separation. Pairs in the same row on the sensor, such as LED~0 \& 1 and LED~0 \& 3, show the lowest inconsistency in this test. Since one row is read at a time, triggering LEDs in a row at different times by applying base delay will exacerbate the inconsistency. However, adding base delay to the LEDs which are not in the same row significantly reduces the relative latency, such as LED~0 \& 4 and LED~0 \& 7. Although minor deviations arise due to secondary factors such as pixel‑level mismatch or localized bus traffic, the overall trend confirms that spatial proximity is a key determinant of temporal synchronization in multi‑region event capture.

% Building on the observation that readout delays depend on spatial location, we further examine how the relative positions of active regions on the sensor affect temporal consistency. Because the sensor relies on a shared readout bus, the spatial distribution of events can influence arbitration efficiency and introduce location‑dependent delays. Using configuration \textbf{I2} with S‑2 aggregation, we focus on scenarios with exactly two active LEDs and evaluate multiple LED pairings, specifically LED 0 paired with LEDs 1, 3, 4, and 7. These combinations represent increasing spatial separation on the sensor.
% For each pairing, we measure the relative delay between the two LEDs across a range of modulation bit rates and under both near and far conditions. As shown in Fig.~\ref{fig:exp6_3_3_delay_across_pair}, inter‑LED readout delay exhibits a strong correlation with spatial separation. Pairs with minimal physical distance on the sensor, such as LED~0–1 and LED~0–4 in the $2 \times 4$ panel, consistently show the smallest relative delays. As separation increases, the measured delay generally increases as well. Although minor deviations arise due to secondary factors such as pixel‑level mismatch or localized bus traffic, the overall trend confirms that spatial proximity is a key determinant of temporal synchronization in multi‑region event capture.

\subsection{Evaluation Results on the Ultra Low-Latency Data Transmission}

\begin{figure}[b]
    \centering
     \vspace{-4mm}
    \subfigure[32 bytes random data.]{
        \includegraphics[width=0.48\linewidth]{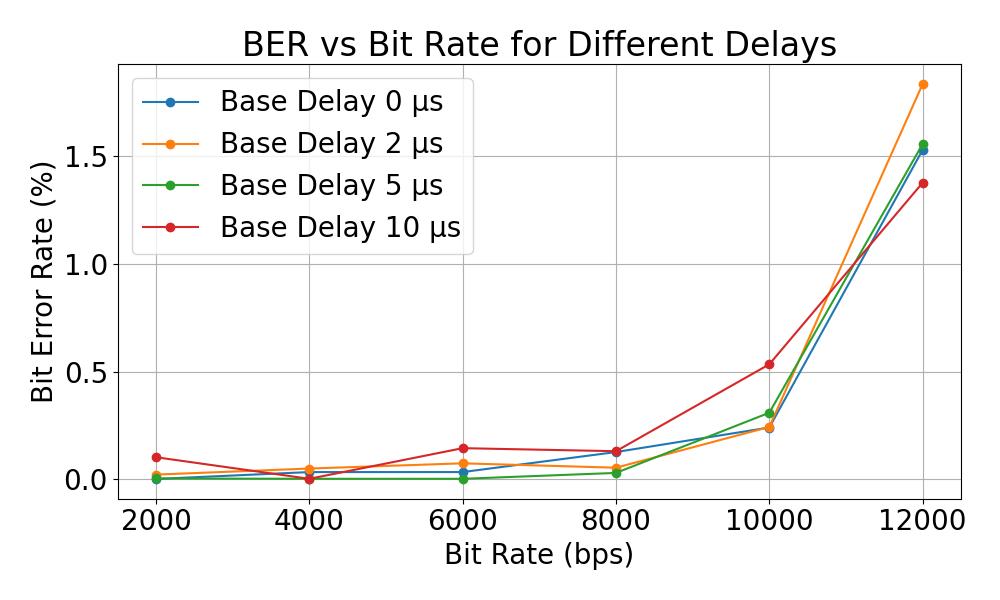}
        \label{fig:exp5_2_random}
    }%
    \hfill
    \subfigure[32 bytes all 1's data.]{
        \includegraphics[width=0.48\linewidth]{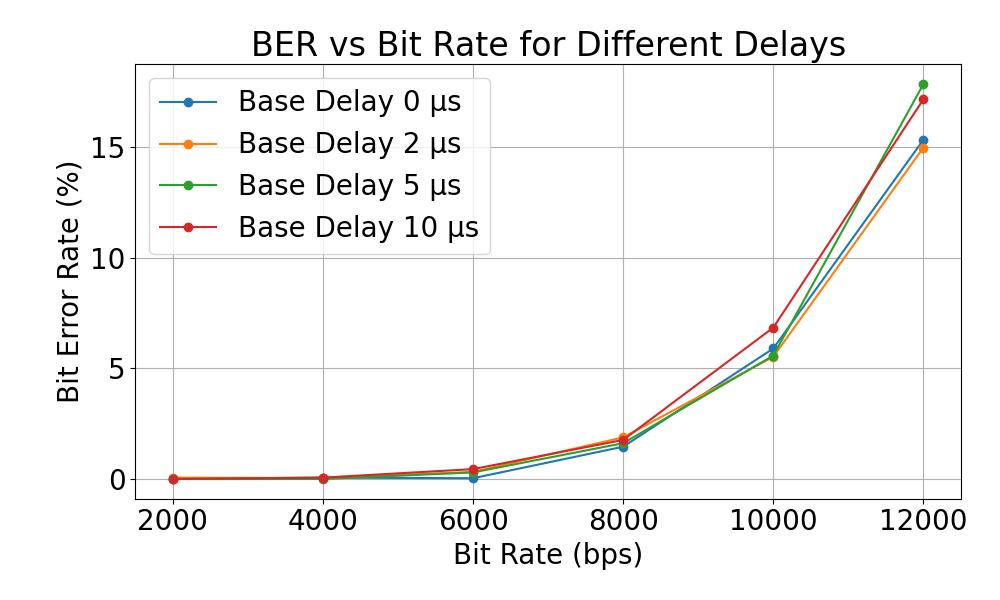}
        \label{fig:exp5_2_worst}
    }
    \vspace{-4mm}
    \caption{Bit error rate vs. transmit bit rate with delays.}
    \label{fig:exp5_2}
\end{figure}

By incorporating the mitigation techniques validated earlier, we evaluate ECO‑COMM under two representative transmission workloads: 32 bytes of random data to model typical application traffic, and 32 bytes of all‑1 data (\texttt{0xFF}) to induce the highest possible event rate. The latter constitutes a deliberate worst‑case scenario, stressing the sensor’s readout bandwidth and processing pipeline.
As shown in Fig.~\ref{fig:exp5_2} and Table~\ref{tab:ber_random_vs_all1}, ECO‑COMM exhibits a well‑balanced reliability–latency trade‑off across a broad range of bit rates. For \emph{random payloads}, performance is particularly strong at moderate rates. At 8{,}000~bps and 10{,}000~bps, the BER remains consistently low, below 0.1\% and 0.5\% respectively, while end‑to‑end latency stays below 10~ms for a full 32‑byte transmission. This operating region represents a favorable balance between reliability and ultra‑low latency, confirming ECO‑COMM’s suitability for time‑critical communication.
As the bit rate increases to 12{,}000~bps, the BER for random data rises modestly to approximately 1.5–1.8\%, reflecting tighter timing margins at higher symbol rates, while latency continues to decrease. This behavior reveals a tunable throughput–reliability trade‑off that can be adapted to application requirements.
For \emph{all‑1 payloads}, ECO‑COMM maintains comparable end‑to‑end latency, indicating that transmission timing is largely insensitive to payload content. Although BER increases more rapidly with bit rate due to reduced intensity transitions, performance degrades smoothly and predictably without abrupt failure. Overall, these results demonstrate that ECO‑COMM is highly effective for transition‑rich data streams and remains robust under adversarial signaling patterns, highlighting its flexibility and practicality for a wide class of ultra‑low‑latency applications.

\begin{table}[t]
\centering
\small
\setlength{\tabcolsep}{3pt}
\caption{BER, association latency (Assoc.), and end‑to‑end latency (E2E) for 32‑byte transmissions.}
\label{tab:ber_random_vs_all1}
\begin{tabular}{c|c|cccc|c|c}
\hline
\textbf{Bps} & \textbf{Hold}
& \multicolumn{4}{c|}{\textbf{BER (\%) @ Delay ($\mu$s)}}
& \textbf{Assoc.}
& \textbf{E2E} \\
 & ($\mu$s)
& 0 & 2 & 5 & 10
& ($\mu$s)
& (ms) \\
\hline
\multicolumn{8}{c}{\textbf{Random Bytes}} \\
\hline
2000  & 0   & 0.0 & 0.0 & 0.0 & 0.1 & 13.4 & 32.05 \\
4000  & 250 & 0.0 & 0.1 & 0.0 & 0.0 & 13.6 & 16.32 \\
6000  & 332 & 0.0 & 0.1 & 0.0 & 0.1 & 13.7 & 11.07 \\
8000  & 250 & 0.1 & 0.1 & 0.0 & 0.1 & 13.5 & 8.32  \\
10000 & 300 & 0.2 & 0.2 & 0.3 & 0.5 & 13.1 & 6.77  \\
12000 & 332 & 1.5 & 1.8 & 1.6 & 1.4 & 13.3 & 5.73  \\
\hline
\multicolumn{8}{c}{\textbf{All‑1 Data (0xFF)}} \\
\hline
2000  & 0   & 0.0 & 0.1 & 0.0 & 0.0 & 14.2 & 32.04 \\
4000  & 250 & 0.1 & 0.0 & 0.0 & 0.1 & 13.2 & 16.39 \\
6000  & 332 & 0.0 & 0.3 & 0.3 & 0.4 & 14.6 & 11.13 \\
8000  & 250 & 1.5 & 1.9 & 1.6 & 1.8 & 13.9 & 8.37  \\
10000 & 300 & 5.9 & 5.5 & 5.6 & 6.8 & 13.3 & 6.80  \\
12000 & 332 & 15.3 & 15.0 & 17.9 & 17.2 & 13.4 & 5.75 \\
\hline
\end{tabular}
\vspace{-5mm}
\end{table}

\vspace{-2mm}
\section{Conclusion}\label{sec:conclusion}

In this paper, we presented ECO‑COMM, an ultra low‑latency optical communication system based on event‑based vision sensing. We identified key hardware and pipeline‑induced challenges that hinder accurate event recording, including timestamp inconsistency and missed detections, and proposed hardware‑aware mitigation techniques to address them. A fully implemented prototype using an eight‑LED transmitter and an off‑the‑shelf event camera demonstrated 10 $\mu$s-level device association, 100 $\mu$s-level symbol latency, and sub‑10 ms end‑to‑end latency for payloads up to 32 bytes. By explicitly accounting for sensor behavior and leveraging the asynchronous nature of event‑based sensing, ECO‑COMM establishes a practical foundation for ultra‑low‑latency optical communication. 
Rather than replacing RF technologies such as Wi-Fi, Bluetooth, NFC, and RFID, ECO-COMM complements them by enabling ultra-low-latency association and time-critical information exchange, advancing event-camera-based optical communication as a viable component of future low-latency wireless systems.
Although ECO-COMM focuses on a single optical communication link, the underlying event-camera communication paradigm has the potential to be extended to multi-link and multi-device environments. Such extensions could support applications including XR systems, robotic coordination, and distributed sensing, where rapid discovery and lightweight information exchange are important. Realizing a networked ECO-COMM framework would require addressing several open challenges, including transmitter identification and addressing, concurrent transmission scheduling, scalable management of multiple regions of interest, mobility-aware tracking, and higher-layer protocol design for association, coordination, and reliability. We view the current system as a foundational proof of concept that motivates future research in these directions.

% \begin{acks}
% TBA
% \end{acks}

\bibliographystyle{ACM-Reference-Format}
\bibliography{reference}

% %%
% %% If your work has an appendix, this is the place to put it.
% \appendix


\end{document}